\documentclass[prd,aps,onecolumn,floatfix,superscriptaddress,preprintnumbers,nofootinbib]{revtex4-2}

\usepackage{svg}

\usepackage[normalem]{ulem} %to strike the words

\usepackage{graphicx,color}
\usepackage{bm}% bold math
\usepackage{amsmath}
\usepackage{amssymb}    
\usepackage{pifont}
\usepackage{cancel}

\usepackage{standalone}
\usepackage{slashed}
\usepackage{multirow}
\usepackage{hhline}
\usepackage{colortbl}
\usepackage{accents} 
\newcommand{\nn}{\nonumber}

\newcommand{\ot}{\leftarrow}

\renewcommand{\(}{\left(}
\renewcommand{\)}{\right)}
\renewcommand{\[}{\left[}
\renewcommand{\]}{\right]}

\renewcommand{\vec}[1]{\bm{#1}}

\newcommand{\MS}{{\overline{\rm MS}}}

\usepackage{ulem}

\DeclareRobustCommand{\sec}[1]{Sec.~\ref{sec:#1}}
\DeclareRobustCommand{\eqref}[1]{Eq.~(\ref{eq:#1})}

\begin{document} 

\title{Transverse momentum moments}

\date{\today}
\newcommand*{\UCM}{Departamento de F\'isica Te\'orica \& IPARCOS, Universidad Complutense de Madridx, Plaza de Ciencias 1, E-28040 Madrid, Spain}\affiliation{\UCM}
\newcommand*{\PSU}{Division of Science, Penn State University Berks, Reading, Pennsylvania 19610, USA}\affiliation{\PSU}
\newcommand*{\JLAB}{Jefferson Lab, Newport News, VA 23606, USA}\affiliation{\JLAB}

\author{Oscar~del~Rio}\affiliation{\UCM}
\author{Alexei~Prokudin}\affiliation{\PSU}\affiliation{\JLAB}
\author{Ignazio~Scimemi}\affiliation{\UCM}
\author{Alexey~Vladimirov}\affiliation{\UCM}

\begin{abstract}
We establish robust relations between Transverse Momentum Dependent distributions (TMDs) and collinear distributions. We define weighted integrals of TMDs that we call Transverse Momentum Moments (TMMs) and prove that TMMs are equal to collinear distributions evaluated in some minimal subtraction scheme. The conversion to the modified minimal subtraction ($\MS$) scheme, can be done by a calculable factor, which we derive up to three loops for some cases.
We discuss in detail the zeroth, the first, and the second TMMs and provide phenomenological results for them based on the current extractions of TMDs. The results of this paper open new avenues for theoretical and phenomenological investigation of the three-dimensional and collinear hadron structures.  
\end{abstract} 

\preprint{IPARCOS-UCM-2024-007, JLAB-THY-24-3989}

\allowdisplaybreaks

\maketitle 
\section{Introduction}
\label{sec:introduction}
Parton distributions are of utmost importance in modern physics, serving a dual purpose. Firstly, they provide information about the internal hadron structure and the nonperturbative dynamics of Quantum Chromodynamics (QCD)~\cite{Gross:2022hyw,Achenbach:2023pba}. Secondly, they allow for nonperturbative QCD effects to be systematically included in theoretical estimates of collider observables~\cite{Lai:2010vv} within the framework of QCD factorization theorems. Important examples of factorization theorems are the collinear and Transverse Momentum Dependent (TMD) factorization theorems~\cite{Collins:2011zzd}. The former approximates the momentum of the active parton to be collinear to the large momentum of the parent hadron and it defines parton distribution functions (PDFs) that encode the information on the momentum fraction, Bjorken $x$, of the hadron's momentum carried by the parton. PDFs are the main source of knowledge regarding collinear (1D) hadron structure and they are determined in global QCD analyses of the experimental measurements. TMD factorization goes further, incorporating the parton's transverse momentum, $\vec{k}_T$, along with the momentum fraction $x$, and it defines Transverse Momentum Dependent parton distribution functions (collectively called TMDs). TMDs provide plentiful information on the three-dimensional (3D) hadron internal structure in the momentum space. Exploration of the 1D and the 3D hadron structures is a cornerstone of existing (LHC~\cite{Begel:2022kwp, Amoroso:2022eow}, Belle~\cite{Accardi:2022oog, Belle-II:2022cgf}, RHIC~\cite{Aschenauer:2015eha}, JLab 12~\cite{Dudek:2012vr}) and future experimental facilities, such as the Electron-Ion Collider~\cite{Boer:2011fh, Accardi:2012qut, AbdulKhalek:2021gbh}. They are also objects of intensive theoretical studies~\cite{Gross:2022hyw,Achenbach:2023pba}, including global phenomenological analyses~\cite{Bertone:2019nxa, Scimemi:2019cmh, Bacchetta:2019sam, Camarda:2019zyx, Ebert:2020dfc, Camarda:2021ict, Bacchetta:2022awv} and higher loop QCD calculations. The most recent TMD analyses include perturbative information up to approximate next-to-next-to-next-to-next-to-leading logarithms (N4LL)~\cite{Neumann:2022lft,Moos:2023yfa}, which is a remarkable success for TMD phenomenology.

It is known that 1D and 3D structures are intimately related. Experimental data from the same processes are used to extract both 3D and 1D structures. For instance, totally inclusive Drell-Yan measurements or high transverse momentum data are used to extract collinear densities while the transverse momentum dependent Drell-Yan cross sections at small values of transverse momentum are relevant to 3D structure.
Intuitively, 3D TMDs can be reduced to 1D PDFs by ``integrating out'' the parton's transverse momentum $\vec k_T$~\cite{Mulders:1995dh,Boer:1997nt}.  The integral relations are explicit in models with parton model approximation and are used widely in phenomenology of TMDs, i.e. Refs.~\cite{Gamberg:2022kdb,Cammarota:2020qcw,Boglione:2018dqd,Anselmino:2013lza}. Integral moments of TMDs were explored in relation to the QCD energy-momentum tensor in Ref.~\cite{Lorce:2015lna}.
In the framework of QCD resummation and small $x$,  the subject was discussed in Refs.~\cite{Catani:1990eg,Catani:1990xk,Catani:1993rn,Catani:1994sq}.
The relationship between the modified minimal subtraction ($\MS$) scheme and other schemes was explored in the resummation framework in Refs.~\cite{Catani:1993ww,Ciafaloni:2005cg,Ciafaloni:2006yk}.

The theoretical relations between TMD and PDF are intricate, and they can be derived in the form of an asymptotic expansion using the Operator Product Expansion (OPE)~\cite{Luo:2020epw, Ebert:2020yqt, Ebert:2020qef, Moos:2020wvd}. Despite being theoretically clear, this approach cannot be applied phenomenologically to relate TMDs and PDFs in a straightforward manner. The main reason is the mismatch of the scale dependence of PDFs and TMDs.  As a consequence of the factorization theorem, TMDs depend on two scales: the ultraviolet (UV) renormalization scale and the renormalization scale of the rapidity divergence, whereas PDFs depend solely on the UV renormalization scale. The corresponding evolution equations have a different analytical structure: a pair of differential Collins-Soper equations diagonal in flavor space with a nonperturbative Collins-Soper (CS) kernel~\cite{Collins:1981uk, Collins:1981va} vs. nondiagonal in flavor space integrodifferential Dokshitzer–Gribov–Lipatov–Altarelli–Parisi (DGLAP) equations~\cite{Gribov:1972ri, Altarelli:1977zs} or DGLAP-type ones \cite{Kang:2008ey,Vogelsang:2009pj,Braun:2009mi,Braun:2009vc} for higher twist distributions. Another reason is the divergence of the integral of TMDs over $\vec k_T$, which is equivalent to the UV divergence of PDFs and should be regularized~\cite{Ebert:2022cku,Gonzalez-Hernandez:2022ifv}. The relation between 3D and 1D structures should be devised in such a way so that the corresponding quantities obtained from 3D densities obey the same evolution equations as their 1D counterparts, while simultaneously exhibiting no dependence on the Collins-Soper kernel.

In this work, we establish a rigorous relationship between 1D and 3D structures that can be readily applied in phenomenology. We define the Transverse Momentum Moments (TMMs) which are weighted integrals of TMDs in the transverse momentum with an upper cutoff. We prove that TMMs are equivalent to the collinear matrix elements computed in a minimal subtraction scheme (we call it TMD scheme), provided that the perturbative scale is high enough to neglect contributions from the power-suppressed terms. The difference between the TMD scheme and $\MS$ scheme emerges at the next-to-leading order (NLO) in the strong coupling, $\alpha_s$, and can be expressed as a calculable factor. We present specific expressions for various cases in this paper.

Specifically, TMMs of order $n$ are defined as the momentum integral with a weight $\vec k_T^n$, and the procedures presented in this work can be defined for any $n\ge 0$. 
However, the estimates for values of $n\ge 2$ are spoiled by the need for subtraction of power divergences. For this reason, we consider only the lowest-power weights ($n=0,1,2$) as they offer the most interesting phenomenological information:
\begin{enumerate}
\item The zeroth TMM, also explored in Refs.~\cite{Ebert:2022cku,Gonzalez-Hernandez:2023iso}, relates TMDs and collinear parton distribution functions (twist-two PDFs). We provide the next-to-next-to-next-to-leading order (N$^3$LO) expression for the matching between the TMD and $\MS$ schemes. The derived relations are verified numerically using unpolarized TMDs from the recent N4LL extraction ART23~\cite{Moos:2023yfa,artemide}, demonstrating the consistency of the method.
\item The first TMM provides information about the distribution of the average transverse momentum shift of partons~\cite{Boer:1997nt,Boer:1997bw,Burkardt:2003yg,Gamberg:2017jha}, and it is expressed via collinear distributions of twist three. Unlike the zeroth TMM, the transformation to the $\MS$ scheme is not applicable here due to the loss of information upon integration. However, the first TMM yields important information. As an illustrative example, we consider the extraction of the Sivers function at N$^3$LO~\cite{Bury:2020vhj, Bury:2021sue} and compute the Qiu-Sterman functions~\cite{Qiu:1991pp}, along with the average transverse momentum shifts of quarks in a transversely polarized proton.
\item The second moment is related to the average absolute value of transverse momentum and the quadrupole distribution, and it is generally described by collinear operators of twist four. Reduction to the TMD scheme requires a subtraction procedure of power divergences, which we devise for the unpolarized case. Results are substantiated phenomenologically using the data from ART23~\cite{Moos:2023yfa}.
\end{enumerate}

The question of scale selection is one of the central in the phenomenology of TMDs. It is also central in the derivations present in this paper because the scaling of TMDs depends on the nonperturbative Collins-Soper kernel. The Collins-Soper kernel does not play a role in any collinear object, and thus the equivalence between TMMs and collinear distributions can be reached only if TMMs are not dependent on the Collins-Soper kernel. We show that it can be achieved in the $\zeta$ prescription \cite{Scimemi:2018xaf} or for a specific selection of general scales in a generic TMD formalism. Notably, the expressions are significantly simpler for the  $\zeta$ prescription, and therefore, we first present the derivation for the $\zeta$ prescription and subsequently explore the case of general scales.

The results of this paper open new avenues for exploration of the three-dimensional structure of the nucleon and in establishing rigorous connections between collinear and transverse momentum dependent observables. They have the potential to become instrumental in phenomenological studies and to serve as benchmarks for comparison with lattice QCD calculations.

The paper is organized as follows: we define the TMMs and the corresponding collinear distributions in~\sec{TMM}. Then, in~\sec{parametrization}, we recall the standard parametrization and properties of TMDs, and in~\sec{evolution} we review the critical elements of TMD evolution and the $\zeta$ prescription. The zeroth, the first, and the second TMMs are considered in Secs.~\ref{sec:0th}, \ref{sec:1st}, and \ref{sec:2nd} correspondingly. The Appendix contains  formulas of the perturbative ingredients for the $\zeta$ prescription.

\section{Transverse Momentum Moments \label{sec:TMM}}

Let us consider a generic quark TMD defined as the following matrix element, see e.g. Ref.~\cite{Boussarie:2023izj}:
\begin{eqnarray}\label{def:F(x,b)}
\widetilde{F}^{[\Gamma]}(x,b)&=& \int \frac{dz}{2\pi} e^{-ixzp^+}\langle p,s|\bar q(zn+b)\mathcal{W}^\dagger_\infty\frac{\Gamma}{2}\mathcal{W}_\infty q(0)|p,s\rangle,
\end{eqnarray}
where $\mathcal{W}_\infty$ represents the gauge link from the quark position to the light cone infinity, the Dirac matrix $\Gamma$ selects the polarization of quarks, $p^+$ denotes the plus component of the momentum  $p$ of the nucleon, $x$ is the momentum fraction carried by the quark, and $b$ is the transverse displacement of the quark fields. The direction of the Wilson line $\mathcal{W}_\infty$ is determined by the physical process~\cite{Collins:2002kn}. TMDs in the momentum space are obtained through the two-dimensional Fourier transform:
\begin{eqnarray}\label{eq:ft}
F^{[\Gamma]}(x,k_T)=\int \frac{d^2\vec b}{(2\pi)^2}e^{i(\vec b\vec k_T)}\widetilde{F}^{[\Gamma]}(x,b),
\end{eqnarray}
where bold symbols denote Euclidean two-dimensional vectors in the transverse plane ($\vec b^2=-b^2>0$). The inverse Fourier transform of Eq.~(\ref{eq:ft}) reads as
\begin{eqnarray}\label{eq:inverse ft}
\widetilde{F}^{[\Gamma]}(x,b)=\int {d^2\vec k_T} e^{-i(\vec b\vec k_T)}F^{[\Gamma]}(x,k_T).
\end{eqnarray}
In this paper, we adopt a notation convention wherein functions with or without tilde denote conjugated functions in $b$ or $k_T$ spaces correspondingly.

The collinear matrix elements of interest are defined as 
\begin{eqnarray}\label{def:Mn-operator}
\mathbb{M}_{\mu_1...\mu_n}^{[\Gamma]}(x)
&=& i^n\int \frac{dz}{2\pi} e^{-ixzp^+}\langle p,s|\bar q(zn)\mathcal{W}_\infty^\dagger \overleftarrow{D}_{\mu_1}...\overleftarrow{D}_{\mu_n} \frac{\Gamma}{2}\mathcal{W}_\infty q(0)|p,s\rangle,
\end{eqnarray}
where $D$ is the covariant derivative, $\overleftarrow{D}_\mu\equiv \overleftarrow{\partial}_\mu + i g A_\mu$. Notice that, the gauge field $A_\mu$ is located at the light cone infinity $\pm n\infty$, and therefore the covariant derivative reduces to the partial derivative in regular gauges. In the case of $n=0$, the matrix element in Eq.~(\ref{def:Mn-operator}) coincides with the definition of twist-2  collinear PDFs.

In the limit of a non interacting theory, that is, in the parton model, one  has
\begin{eqnarray}\label{def:Mn-bare}
\mathbb{M}_{\mu_1...\mu_n}^{[\Gamma]}(x)%\stackrel{\rm parton \; model}
\,\xrightarrow[]{\rm parton\; model}\,i^n \lim_{b\to0}\frac{\partial^n}{\partial \vec b^{\mu_1}...\partial \vec  b^{\mu_n}}\widetilde{F}^{[\Gamma]}(x,b)=\int d^2\vec k_T \, \vec k_{T\mu_1}...\vec k_{T\mu_n} F^{[\Gamma]}(x,k_T),
\end{eqnarray}
where all vectors are Euclidean. If one interprets a TMD as a parton distribution with momentum $x p^\mu+ k_T^\mu$ within a hadron of momentum  $p^\mu$, then the TMMs can be regarded as the parton distribution of the averaged transverse momenta. For instance, according to this interpretation, $-g^{\mu\nu}_T\mathbb{M}_{\mu\nu}$ represents the average $\langle \vec k_T^2 \rangle$ distribution of a parton in a hadron.

Accounting for interactions requires a renormalization procedure, which is however different for the left- and right-hand sides of Eq.~(\ref{def:Mn-bare}). 
The renormalization of the TMD operator needs both the ultra-violet (UV) and rapidity renormalization scale, respectively $\mu$ and $\zeta$, and as a result, TMDs become dependent on them $F(x,k_T; \mu, \zeta)$. The renormalization of Eq.~(\ref{def:Mn-operator}) is purely UV and it proceeds with a single scale, and therefore Eq.~(\ref{def:Mn-operator}) acquires the scale dependence $\mathbb{M}_{\mu_1...\mu_n}^{[\Gamma]}(x,\mu)$. As a result, the scale dependence of both sides is determined by different classes of evolution equations. The evolution of TMD has a double-logarithmic nature~\cite{Collins:2011zzd,Becher:2010tm,Echevarria:2011epo,Chiu:2012ir,Ebert:2021jhy,Vladimirov:2021hdn}, while matrix elements in Eq.~(\ref{def:Mn-operator}) evolve by integrodifferential DGLAP-type equations~\cite{Gribov:1972ri, Altarelli:1977zs,Kang:2008ey,Vogelsang:2009pj,Braun:2009mi,Braun:2009vc}.

The TMDs are the fundamental matrix elements of the factorized cross section and on top of this, they can also be refactorized in the limit $b\to0$. This second factorization is the result of the OPE which was studied thoroughly~\cite{Aybat:2011zv, Bacchetta:2013pqa, Echevarria:2015usa, Echevarria:2016scs, Gutierrez-Reyes:2017glx, Gutierrez-Reyes:2018iod, Gutierrez-Reyes:2019rug, Luo:2019szz, Luo:2019hmp, Scimemi:2019gge, Ebert:2020yqt, Ebert:2020qef, Luo:2020epw, Rein:2022odl}.  Conventionally one uses an auxilliary scale $\mu_{\text{OPE}} \propto 1/b$ in the OPE for TMDs to minimize logarithmic corrections.  The OPE allows one to determine the asymptotic behavior of TMDs in terms of collinear distributions, but it cannot be easily inverted. The main complication comes from different classes of evolution equations. 

In this paper, we consider weighted integrals (\ref{def:Mn-bare}) with a momentum cutoff $|k_T|<\mu$. We show that they are equivalent to the corresponding collinear distributions. The critical element of our construction is the selection of scales for TMDs. It must be done such that the nonperturbative Collins-Soper kernel is eliminated as the collinear matrix elements do not depend on it. We identify two cases where this elimination can be achieved~\footnote{Other constructions satisfying these criteria may exist.}:
\begin{enumerate}
\item The TMD is evaluated using the $\zeta$ prescription \cite{Scimemi:2018xaf}. In this case, TMMs are defined as
\begin{eqnarray}\label{def:TMM}
\mathcal{M}_{\nu_1...\nu_n}^{[\Gamma]}(x,\mu)\equiv
\int^\mu d^2\vec k_T \, \vec k_{T\nu_1}...\vec k_{T\nu_n} F^{[\Gamma]}(x,k_T),
\end{eqnarray}
where the TMD on the right-hand side is the optimal TMD (see \sec{evolution} for details). It is defined at the point of vanishing Collins-Soper kernel, and thus any TMM does not depend on the Collins-Soper kernel.
\item The TMD is evaluated with all involved scales equal to $\mu$, i.e.
\begin{eqnarray}\label{def:TMM2}
\accentset{\ast}{\mathcal{M}}_{\nu_1...\nu_n}^{[\Gamma]}(x,\mu)\equiv
\int^\mu d^2\vec k_T \, \vec k_{T\nu_1}...\vec k_{T\nu_n} F^{[\Gamma]}(x,k_T;\mu,\mu^2).
\end{eqnarray}
In this case, the elimination of the Collins-Soper kernel is not by construction but it happens due to the properties of the OPE and the TMM integral at $n=0,1,2$. 
\end{enumerate}

We call these integrals in Eqs.~(\ref{def:TMM}) and (\ref{def:TMM2}) TMMs. In what follows we show that (and similarly for $\accentset{\ast}{\mathcal{M}}$)
\begin{eqnarray}
\label{eq:Mmu}
\mathcal{M}_{\nu_1...\nu_n}^{[\Gamma]}(x,\mu) = \mathbb{M}_{\nu_1...\nu_n}^{[\Gamma]}(x,\mu)
+\mathcal{O}(\mu^{-2}),
\end{eqnarray}
where we assume that the cutoff scale $\mu$ is sufficiently large, to neglect $\mathcal{O}(\mu^{-2})$ corrections. We prove that the TMMs obtained with Eqs.~(\ref{def:TMM},\ref{def:TMM2}) (after the appropriate subtractions) coincide with the collinear quantities from Eq.~(\ref{def:Mn-operator}) computed in some minimal subtraction scheme, that we call the TMD scheme for $\mathcal{M}$ and TMD2 scheme for $\accentset{\ast}{\mathcal{M}}$. It means that TMM obeys the same evolution as the corresponding collinear matrix element, with the kernel that differs from the one in $\MS$ at order $\alpha_s^2$ (i.e. at NLO). In some cases, it is possible to introduce finite renormalization constants to match the TMD scheme, TMD2 scheme, and the $\overline{\text{MS}}$ scheme. Note, that the cutoff parameter $\mu$ in Eqs.~(\ref{def:TMM}) and (\ref{def:TMM2}) is the renormalization scale for the collinear distribution.

Importantly, the numerical values of (\ref{def:TMM}) and (\ref{def:TMM2}) are not the same, despite both being equivalent to the same collinear matrix element. TMMs (\ref{def:TMM}) and (\ref{def:TMM2}) correspond to collinear matrix elements evaluated in different schemes, which can be reduced to the $\MS$ scheme by different factors (see \sec{0th}). In other words,
\begin{eqnarray}
\mathcal{M}_{\nu_1...\nu_n}^{[\Gamma]}(x,\mu)/\accentset{\ast}{\mathcal{M}}_{\nu_1...\nu_n}^{[\Gamma]}(x,\mu)=1+\mathcal{O}(\alpha_s).
\end{eqnarray}

Detailed discussions on the relationship between the zeroth moment of unpolarized TMDs or Sivers TMD and the corresponding collinear functions can be found in Refs.~\cite{Bacchetta:2013pqa, Gamberg:2017jha, Qiu:2020oqr, Ebert:2022cku, Gonzalez-Hernandez:2023iso, Bury:2021sue}. Higher moments are more intricate as they are associated with power corrections in the OPE.

\section{Parametrization of TMDs in position and momentum space \label{sec:parametrization}}

The standard parametrization \cite{Boussarie:2023izj} of the TMD matrix element in Eq.~(\ref{def:F(x,b)}) reads as
\begin{eqnarray}\nn
\widetilde{F}^{[\gamma^+]}(x,b)&=&\widetilde{f}_1(x,b)+i\epsilon^{\mu\nu}_T b_\mu s_{T\nu}M \widetilde{f}_{1T}^\perp(x,b),
\\\label{def:parametrization-b}
\widetilde{F}^{[\gamma^+\gamma^5]}(x,b)&=&\lambda \widetilde{g}_{1}(x,b)+i(b \cdot s_T)M \widetilde{g}^\perp_{1T}(x,b),
\\\nn
\widetilde{F}^{[i\sigma^{\alpha+}\gamma^5]}(x,b)&=&s_T^\alpha \widetilde{h}_{1}(x,b)-i\lambda b^\alpha M \widetilde{h}_{1L}^\perp(x,b) \label{eq:b-parametrization}
\\\nn && +i\epsilon_T^{\alpha\mu}b_\mu M \widetilde{h}_1^\perp(x,b)+\frac{M^2 {\bm b^2}}{2}\(\frac{g_T^{\alpha\mu}}{2}+\frac{b^\alpha b^\mu}{\bm b^2}\)s_{T\mu}\widetilde{h}_{1T}^\perp(x,b),
\end{eqnarray}
where $\epsilon_T^{\mu\nu}=\epsilon^{-+\mu\nu}=\epsilon^{30\mu\nu}$ with $\epsilon_T^{12}=+1$, $g_T^{\mu\nu}=g^{\mu\nu}-n^\mu \bar n^\nu-\bar n^\mu n^\nu$, $\lambda$ and $s_T$ are the longitudinal and transverse components of the spin vector. The mass parameter $M$  is a typical nonperturbative scale, often chosen to be the mass of the hadron. It is worth noting that a TMD depends solely on the absolute value of the transverse coordinate $b$. In momentum space, the parametrization is given by
\begin{eqnarray}
\nn
F^{[\gamma^+]}(x,k_T)&=&f_1(x,k_T)-\epsilon^{\mu\nu}_T \frac{k_{T\mu} s_{T\nu}}{M} f_{1T}^\perp(x,k_T),
\\\label{def:parametrization-kT}
F^{[\gamma^+\gamma^5]}(x,k_T)&=&\lambda g_{1}(x,k_T)-\frac{(k_T \cdot s_T)}{M} g^\perp_{1T}(x,k_T),
\\\nn
F^{[i\sigma^{\alpha+}\gamma^5]}(x,k_T)&=&s_T^\alpha h_{1}(x,k_T)+\frac{\lambda k_T^\alpha}{M} h_{1L}^\perp(x,k_T)
\\\nn && -\frac{\epsilon_T^{\alpha\mu}k_{T\mu}}{M}h_1^\perp(x,k_T)-\frac{\bm k_T^2}{M^2}
\(\frac{g_T^{\alpha\mu}}{2}+\frac{k_T^\alpha k_T^\mu}{\bm k_T^2}\)s_{T\mu}h_{1T}^\perp(x,k_T).
\end{eqnarray}
The  TMDs in each space are related to each other by the Hankel transform
\begin{eqnarray}\label{def:Jn}
F(x,k_T)=\frac{M^{2n}}{ n!}\int_0^\infty \frac{db b}{2\pi} \left( \frac{b}{k_T} \right)^n  J_n(bk_T)\widetilde F^{(n)}(x,b;\mu,\zeta),
\end{eqnarray}
which is obtained by the angular integration in the Fourier transform of Eq.~(\ref{def:F(x,b)}), as $b$ and $k_T$ are absolute values of the transverse coordinate and momentum vectors. In Eq.~(\ref{def:Jn}) we introduce a superscript $(n)$ and explicitly apply the following relation~\cite{Boer:2011xd} for TMDs in $b$ space:
\begin{align} \label{eq:TMD_bt_derivative}
 \widetilde{F}^{(n)}(x, b_T; \mu, \zeta) &
 \equiv n! \left(\frac{-1}{M^2 b} \partial_{b} \right)^n \widetilde{F}(x, b;\mu, \zeta)
%\nn\\&
 = \frac{2\pi\, n!}{(b M)^n} \int_0^\infty d k_T \, k_T \left(\frac{k_T}{M}\right)^n J_n(b k_T) \, F(x, k_T; \mu, \zeta)
\,.\end{align}
Note that for real functions $f$, this Hankel transform is also real, and $\tilde f^{(n)}$ possesses the same mass dimension for all~$n$. Therefore, the correspondence between the functions in Eq.~(\ref{def:parametrization-b}) and Eq.~(\ref{eq:TMD_bt_derivative}) is
\begin{align} \nn
\widetilde f_1(x,b)&\equiv \widetilde{f}_1^{(0)} (x,b),
& f_{1T}^\perp (x,b)&\equiv \widetilde{f}_{1T}^{\perp (1)} (x,b),
\\
\label{eq:nindex}  \widetilde g_1(x,b)&\equiv\widetilde{g}_1^{(0)} (x,b),
& \widetilde g_{1T}^\perp(x,b)&\equiv\widetilde{g}_{1T}^{\perp (1)} (x,b),
\\
\nn \widetilde h_1(x,b)&\equiv\widetilde{h}_1^{(0)}(x,b),
& \widetilde h_{1L}^\perp(x,b)&\equiv\widetilde{h}_{1L}^{\perp (1)} (x,b),
\\
\nn \widetilde h_1^\perp(x,b)&\equiv\widetilde{h}_1^{\perp (1)} (x,b),
& \widetilde h_{1T}^\perp(x,b)&\equiv\widetilde{h}_{1T}^{\perp (2)} (x,b).
\end{align}
Formally one has~\cite{Boer:2011xd}
\begin{align}\label{eq:nth-moment}
   \lim_{b\to 0}\widetilde f^{(n)}(x,b) = \int d^2 \vec k_T \left(\frac{\vec k_T^2}{2 M^2}\right)^n f(x, k_T) \equiv f^{(n)} (x),
\end{align}
where $f^{(n)}(x)$ is often referred to as the $n$-th moment of the TMD.

The superscripts $(0),(1),(2)$ in Eqs.~(\ref{eq:nindex}) help to keep track of the asymptotic behavior of the TMDs and they are important for the discussion of TMMs in the following sections. The large-$k_T$ asymptotic of the TMDs are power like and in $k_T$ space one has 
\begin{eqnarray}\label{eq:large-kt}
f(x, k_T)\propto \frac{M^{2m}}{(k_T^2)^{m+1}},
\end{eqnarray}
accompanied by powers of logarithms $\ln(k_T)$, where $f$ represents one of the TMDs from the right-hand side of Eq.~(\ref{def:parametrization-kT}), and $m$ is the corresponding superscript of $\widetilde f^{(m)}(x,b)$ from Eqs.~(\ref{eq:nindex}).

\section{Evolution of TMDs and $\zeta$ prescription}
\label{sec:evolution}

The treatment of evolution scales plays a central role in the derivations of relations between 3D and 1D structures. In this section, we recap the main elements of TMD evolution and the $\zeta$ prescription. The detailed definition and derivations can be found in Refs.~\cite{Scimemi:2018xaf, Vladimirov:2019bfa}. For the reader's convenience, we collect several useful formulas in the Appendix.

All TMDs depend on two renormalization scales, $\mu$, the UV evolution scale, and $\zeta$, the rapidity evolution scale. In the position space, the dependence on $(\mu,\zeta)$ is governed by two evolution equations~\cite{Collins:2011zzd},
\begin{eqnarray}
\label{evol:UV}
\mu^2\frac{d }{d\mu^2}\widetilde{F}(x,b;\mu,\zeta)&=&\frac{\gamma_F(\mu,\zeta)}{2} \widetilde{F}(x,b;\mu,\zeta),
\\
\label{evol:rapidity}
\zeta \frac{\partial}{\partial \zeta} \widetilde{F}(x,b;\mu,\zeta) &=& -\mathcal{D}(b,\mu)\widetilde{F}(x,b;\mu,\zeta),
\end{eqnarray}
where $\widetilde{F}$ represents any TMD in $b$ space from the lhs of Eq.~(\ref{def:parametrization-b}), and $\mathcal{D}$ is the Collins-Soper kernel.\footnote{$\mathcal{D}= -\tilde{K}/2$, where $\tilde{K}$ is the Collins-Soper kernel in the notation of Ref~\cite{Collins:2011zzd}. It is a universal object for all quarks. For gluons, the Collins-Soper kernel is different, and we do not explicitly consider gluons in this paper; hence we do not assign the flavor index to it.} The TMD anomalous dimension $\gamma_F$ has the following form:
\begin{eqnarray}\label{def:gammaF}
\gamma_F(\mu,\zeta) \equiv {\Gamma_{\text{cusp}}(\mu)}\ln\(\frac{\mu^2}{\zeta}\)-{\gamma_V(\mu)}.
\end{eqnarray}
Here, $\Gamma_{\text{cusp}}$ is the cusp-anomalous dimension, and $\gamma_V$ is the vector anomalous dimension. The coefficients of the corresponding perturbative series are denoted as
\begin{eqnarray}\label{eq:Gamma}
\Gamma_{\text{cusp}}(\mu)&=&\sum_{n=0}^\infty \alpha_s^{n+1}(\mu)\Gamma_n,
\\\label{eq:gammaV}
\gamma_{V}(\mu)&=&\sum_{n=1}^\infty \alpha_s^{n}(\mu)\gamma_n,
\end{eqnarray}
with $\alpha_s(\mu)$ being the QCD coupling constant. The Collins-Soper kernel is a nonperturbative function that satisfies the following evolution equation:
\begin{eqnarray}\label{CS-equation}
\mu^2 \frac{d}{d\mu^2}\mathcal{D}(b,\mu)=\frac{\Gamma_{\text{cusp}}(\mu)}{2}.
\end{eqnarray}
In the small-$b$ regime, the Collins-Soper kernel can be computed perturbatively. The corresponding expression reads as, see i.e. \cite{Becher:2010tm,Scimemi:2018xaf},
\begin{eqnarray}\label{eq:cs-expansion}
\mathcal{D}(b,\mu)&=&\sum_{n=0}^\infty \sum_{k=0}^{n}\alpha_s^n \mathbf{L}^k_\mu d^{(n,k)}+\mathcal{O}(b^2),
\end{eqnarray}
where 
\begin{eqnarray}
\mathbf{L}_{\mu}=\ln\(\frac{\mu^2 b^2}{4e^{-2\gamma_E}}\).    
\end{eqnarray}
All coefficients $d^{(n,k)}$ with $k\neq0$ can be expressed via recursive formulas using anomalous dimensions, see for instance \cite{Scimemi:2018xaf}, and $d^{(1,0)}=0$. The power corrections $\mathcal{O}(b^2)$ to Eq.~(\ref{eq:cs-expansion}) are expressed via vacuum matrix elements \cite{Vladimirov:2020umg}, and can be determined through comparison with experimental data \cite{BermudezMartinez:2022ctj} or by utilizing lattice QCD calculations \cite{Shu:2023cot, Avkhadiev:2023poz}.

The fact that the evolution of TMDs is determined by a system of evolution equations~(\ref{evol:UV},\ref{evol:rapidity}) leads to several consequences. Firstly, Eqs.~(\ref{CS-equation}) and (\ref{def:gammaF}) guarantee the existence of a solution for the system (\ref{evol:UV}), (\ref{evol:rapidity}) which reads as 
\begin{eqnarray}\label{evol:solution}
F(x,b;\mu,\zeta)=\exp\Big[\int_P \(\gamma_F(\mu',\zeta')\frac{d\mu'}{\mu'}-\mathcal{D}(b,\mu')\frac{d\zeta'}{\zeta'}\)\Big]F(x,b;\mu_0,\zeta_0),
\end{eqnarray}
where $P$ is any path in the ($\mu$,$\zeta$)-plane connecting points $(\mu,\zeta)$ and $(\mu_0,\zeta_0)$.
Secondly, there exist curves in the $(\mu,\zeta)$ space along which TMDs do not evolve~\cite{Scimemi:2018xaf, Vladimirov:2019bfa}. These are the equipotential lines of the two-dimensional vector field ${\bf E} \equiv (\gamma_F(\mu,\zeta)/2, -\mathcal{D}(b,\mu))$~\cite{Scimemi:2018xaf}. Consequently, the selection of a particular scale for a TMD can be equivalently replaced by the selection of a particular equipotential line. The equipotential line $(\mu,\zeta_\mu(b))$ is determined by the following equation~\cite{Vladimirov:2019bfa}:
\begin{eqnarray}\label{zeta-line}
\Gamma_{\text{cusp}}(\mu)\ln\(\frac{\mu^2}{\zeta_\mu(b)}\)-\gamma_V(\mu)=2\mathcal{D}(b,\mu)\frac{d\ln \zeta_\mu(b)}{d\ln \mu^2}.
\end{eqnarray}
Importantly, this equation and the Collins-Soper kernel within it are applicable across all values of $b$, including the nonperturbative large-$b$ region. In this sense, the value of $\zeta_\mu$ is a functional of $\mathcal{D}$, denoted as $\zeta_\mu(b)\equiv \zeta_\mu(\mathcal{D}(b))$.

Therefore, the $\zeta$ prescription consists in defining the TMD on the equipotential line $(\mu,\zeta_\mu(b))$. By definition, such a TMD, which we call TMD in $\zeta$ prescription, is scale invariant, satisfying
\begin{equation}\label{eq:tmd-in-zeta-prescription}
\mu^2\frac{d}{d\mu^2}F(x,b;\mu,\zeta_\mu(b))=0.
\end{equation}
To obtain the TMD at experimentally observed scales $(\mu,\zeta)$, one evolves the TMD from $(\mu_0,\zeta_{\mu_0})$ to $(\mu,\zeta)$. Notably, any convenient value of $\mu_0$ may be chosen, as the TMD in $\zeta$ prescription $F(x,b;\mu,\zeta_\mu(b))$ remains independent of it. A straightforward choice is $\mu_0=\mu$, and the simplest evolution path is a straight line from $\zeta_0=\zeta(\mu)$ to $\zeta=\mu^2$. In this case, the evolution takes the form of a multiplicative factor,
\begin{eqnarray}\label{def:opt->norm}
F(x,b;\mu,\zeta)=\(\frac{\zeta}{\zeta_\mu(b)}\)^{-\mathcal{D}(b,\mu)}F(x,b;\mu,\zeta_\mu(b)).
\end{eqnarray}
This formulation facilitates the computation of the TMD at any chosen set of scales $(\mu,\zeta_\mu(b))$ starting from the TMD $F(x,b;\mu,\zeta_\mu(b))$ in the  $\zeta$ prescription, as given by Eq.~(\ref{eq:tmd-in-zeta-prescription}).

In this approach, the selection of the scale for TMDs is replaced by the selection of an equipotential line. It was demonstrated in Ref.~\cite{Vladimirov:2019bfa} that the optimal selection is the equipotential line passing through the saddle point $(\mu_0,\zeta_0)$ of the evolution field. The saddle point is defined as
\begin{eqnarray}\label{def:saddle}
\mathcal{D}(b,\mu_0)=0,
\end{eqnarray}
\begin{eqnarray}\label{def:saddle1}
\Gamma_{\text{cusp}}(\mu_0)\ln\(\frac{\mu_0^2}{\zeta_0}\)-\gamma_V(\mu_0)=0. 
\end{eqnarray}
This point has the advantage of being uniquely defined and such that the values of $\zeta_\mu$ are finite at all values of $\mu$ (which is not guaranteed for any arbitrary equipotential line defined by Eq.~(\ref{zeta-line})). The TMD defined on the line that passes through the saddle point, Eqs.~(\ref{def:saddle}) and (\ref{def:saddle1}), is called the optimal TMD and is conventionally defined without explicit scales, as discussed in Refs.\cite{Scimemi:2018xaf, Vladimirov:2019bfa}.  Another advantage of the optimal TMD on the equipotential line passing through the saddle point is that the resulting TMD is inherently independent of the Collins-Soper kernel by construction, due to Eq.~(\ref{def:saddle}).

Phenomenological studies that use $\zeta$ prescription proved to be very fruitful and include Refs.~\cite{Scimemi:2018xaf, Vladimirov:2019bfa, Bertone:2019nxa, Vladimirov:2019bfa, Scimemi:2019cmh, Hautmann:2020cyp, Bury:2020vhj, Bury:2021sue, Bury:2022czx, Horstmann:2022xkk}, along with the latest extraction of TMDs from Drell-Yan data performed at an approximate N4LL accuracy~\cite{Moos:2023yfa}.

\section{The zeroth Transverse Momentum Moment}
\label{sec:0th}

The zeroth TMM relates TMDs to the collinear (twist two) parton distribution functions. In this section, we demonstrate that the zeroth TMM corresponds to the collinear PDF and exhibits the correct DGLAP evolution. Specifically, we show that it can be precisely matched to the $\MS$ collinear PDF through a finite renormalization constant.

The zeroth TMM is simply the momentum integral of the TMD given in Eq.(\ref{def:TMM}). Upon substituting the parametrization for particular Dirac structures (\ref{def:parametrization-b}), we obtain
\begin{eqnarray}
\nn
\mathcal{M}^{[\gamma^+]}(x,\mu) = \int^\mu d^2 {\bm k}_T F^{[\gamma^+]}(x,k_T)&=& \int^\mu d^2 {\bm k}_T f_1(x,k_T),
\\\label{def:zeroth-moment}
\mathcal{M}^{[\gamma^+\gamma_5]}(x,\mu) = \int^\mu d^2 {\bm k}_T F^{[\gamma^+\gamma^5]}(x,k_T)&=& \lambda \int^\mu d^2 {\bm k}_T g_{1}(x,k_T),
\\\nn
\mathcal{M}^{[i\sigma^{\alpha+}\gamma^5]}(x,\mu) = \int^\mu d^2 {\bm k}_T F^{[i\sigma^{\alpha+}\gamma^5]}(x,k_T)&=& s_T^\alpha \int^\mu d^2 {\bm k}_T h_{1}(x,k_T) 
\\\nn &&
- \int^\mu d^2 {\bm k}_T \frac{\vec k_T^2}{M^2}
\(\frac{g_T^{\alpha\mu}}{2}+\frac{k_T^\alpha k_T^\mu}{\vec k_T^2}\)s_{T\mu}h_{1T}^\perp(x,k_T),
\end{eqnarray}
The last term contributes as $\sim \mu^{-2}$, because the pretzelocity TMD, $h_{1T}^\perp$, behaves as $k_T^{-6}$ at large $k_T$. Since we are considering the large-$\mu$ regime, we will neglect this term.

To simplify the notations, we introduce operation $\mathcal{G}$, see also Ref.~\cite{Ebert:2022cku}, 
\begin{eqnarray} \label{def:Gnm}
\mathcal{G}_{n,m} [f](x,\mu)&=&\int^\mu d^2\vec k_T \left(\frac{\vec k_T^2}{2 M^{2}}\right)^nf(x,k_T)\;.
\end{eqnarray}
In Eq.~(\ref{def:Gnm}) the first index $n$ can be any integer, while
the second index $m$ is the index of the TMD $\tilde f$ from  Eqs.~(\ref{eq:nindex}) in $b$ space, that corresponds to the TMD $f$. Notice that without the upper cutoff, one recovers the usual $n$-th moment of the TMD, as given by Eq.~(\ref{eq:nth-moment}). 
With this notation and using Eq.~(\ref{eq:large-kt}) we obtain the following properties:
\begin{eqnarray}
\mathcal{G}_{m,m} [f](x,\mu)&\propto&\ln(\mu) \; , \nn 
\\ \label{def:Gn-divergence}
\mathcal{G}_{m+l,m} [f](x,\mu)&\propto&\mu^{2l} \; \text{for } m + l \geq0  \;. 
\end{eqnarray}
That is, for any TMD in  $b$ space of index $m$, the operation $\mathcal{G}_{m,m}$ exhibits logarithmic divergence, the operation $\mathcal{G}_{m+l,m}$ has powerlike divergence of order $l$ if $l>0$, and $\mathcal{G}_{m+l,m}$ is convergent if $l<0$.
In what follows for the zeroth moment, where indices are $(0,0)$, the index $n$ is not a free parameter but is defined by the index $m$ of the TMD. For simplicity, we use a single subscript in the case where $n=m$, so that $\mathcal{G}_{n,n} \equiv \mathcal{G}_{n}$.
Hence, we have 
\begin{align} %\nn
\mathcal{G}_{n} [f](x,\mu)&=\int^\mu d^2\vec k_T \left(\frac{\vec k_T^2}{2 M^{2}}\right)^nf(x,k_T)
\label{def:Gn} = %&=&
\frac{1}{n!}\int_0^\infty db\,\mu\left(\frac{\mu b}{2}\right)^n J_{n+1}(\mu b)\widetilde{f}^{(n)}(x,b),
\end{align}
where  $J_{n+1}$ is the Bessel function of the first kind. In this notation, Eqs.~(\ref{def:zeroth-moment}) turns into
\begin{eqnarray}\label{def:M0[f]}
\mathcal{M}^{[\gamma^+]}(x,\mu)&=&\mathcal{G}_0[f_1](x,\mu),
\\\label{def:M0[g]}
\mathcal{M}^{[\gamma^+\gamma^5]}(x,\mu)&=&s_L \mathcal{G}_0[g_1](x,\mu),
\\\label{def:M0[h]}
\mathcal{M}^{[i\sigma^{\alpha+}\gamma^5]}(x,\mu)&=&s^\alpha_T\mathcal{G}_0[h_1](x,\mu),
\end{eqnarray}
where $s_L$ and $s_T$ are the longitudinal and transverse components of the proton's spin. TMDs $f_1$, $g_1$, and $h_1$ are known as the unpolarized, helicity, and transversity TMDs, respectively.  

At large values of $k_T$ the TMDs $f_1$, $g_1$, and $h_1$ behave as $k_T^{-2}$, see Eq.~(\ref{eq:large-kt}), potentially multiplied by powers of $\ln(k_T)$~\cite{Bacchetta:2008xw}. Therefore, the integrals $\mathcal{G}_0$ diverge logarithmically as $\mu$ becomes large. This is, in essence, the UV divergence associated with the DGLAP evolution of the corresponding collinear distributions. As a result, the zeroth TMMs, Eqs.~(\ref{def:M0[f]})-(\ref{def:M0[h]}), exhibit logarithmic divergences, and these divergences are similar to those appearing in collinear PDFs. In fact, for sufficiently large cut-off scales $\mu$, the following relations between them can be established:
\begin{eqnarray}\label{G0=f}
\mathcal{G}_0[F](x,\mu)=f^{(\text{TMD})}(x,\mu)+\mathcal{O}(\mu^{-2}),
\end{eqnarray}
where $F$ represents the TMDs $f_1$, $g_1$, and $h_1$, and $f^{(\text{TMD})}$ is related to unpolarized ($q(x)$), helicity ($\Delta q(x)$), and transversity ($\delta q(x)$) collinear PDFs, respectively:
\begin{eqnarray}
\mathcal{G}_0[f_1](x,\mu) &=&  q^{(\text{TMD})}(x,\mu)+\mathcal{O}(\mu^{-2}), \nn \\
\mathcal{G}_0[g_1](x,\mu) &=&  \Delta q^{(\text{TMD})}(x,\mu)+\mathcal{O}(\mu^{-2}), \\\nn
\mathcal{G}_0[h_1](x,\mu) &=&  \delta q^{(\text{TMD})}(x,\mu)+\mathcal{O}(\mu^{-2}).
\end{eqnarray}
These relations were considered in Ref.~\cite{Bacchetta:2013pqa} at NLL and then studied in detail in Ref.~\cite{Ebert:2022cku}. The label ``(TMD)" is used to distinguish the functions resulting from the operation $\mathcal{G}_0$ from the collinear functions themselves. We will demonstrate that the resulting functions $f^{(\text{TMD})}$ obey the same DGLAP evolution equations as collinear PDFs. Therefore, the label  ``(TMD)" indicates that the collinear PDF is evaluated in a particular \textit{TMD scheme}, which is a minimal subtraction scheme, but it does not coincide with the $\overline{\text{MS}}$ scheme \cite{tHooft:1973mfk, Bardeen:1978yd}. The transformation between the TMD scheme and the conventional $\overline{\text{MS}}$ scheme can be performed by a finite renormalization constant $Z^{\MS/\text{TMD}}$, which we derive below. The schemes differ also for  TMMs evaluated with the optimal TMDs, Eq.~(\ref{def:TMM}), or with  TMDs at general scales, Eq.~(\ref{def:TMM2}). We consider these cases one by one, starting with the $\zeta$ prescription.

\subsection{Optimal TMDs}
\label{sec:0th:optimal}

To derive Eq.~(\ref{G0=f}) and verify its properties, we exploit the correspondence between the large-$\mu$ asymptotic behavior of Hankel integrals and the small-$b$ asymptotic behavior of the integrand \cite{slonovskii1968asymptotic}. The small-$b$ asymptotic of a TMD can be computed using the OPE. For TMDs $f_1$, $g_1$, and $h_1$, the OPE takes the form
\begin{eqnarray}\label{small-b:position}
\widetilde{F}_f(x,b)=\sum_{f'}\int_x^1 \frac{dy}{y}\widetilde{C}_{f\ot f'}(y,\mu_{\text{OPE}})f_{f'}\(\frac{x}{y},\mu_{\text{OPE}}\)+\mathcal{O}(b^2) \equiv \widetilde{C}\otimes f +\mathcal{O}(b^2),
\end{eqnarray}
where $ \widetilde{C}$ is the coefficient function that depends on $b$ via $\mathbf{L}_{\mathcal{O}}=\ln(\mu^2_{\text{OPE}}\vec b^2/4e^{-2\gamma_E})$. Here we explicitly indicate the flavor labels $f$ of the TMD and introduce the convolution notation ``$\otimes$'' that implies Mellin convolution and summation over flavors $f'$ (quarks, antiquarks, and gluons)
\begin{eqnarray}\label{eq:convolution}
\widetilde{C}\otimes f\equiv\sum_{f'}\int_x^1 \frac{dy}{y}\widetilde{C}_{f\ot f'}\(y,\mu\)f_{f'}\(\frac{x}{y},\mu\) = \sum_{f'}\int_x^1 \frac{dy}{y}\widetilde{C}_{f\ot f'}\(\frac{x}{y},\mu\)f_{f'}\(y,\mu\).
\end{eqnarray}
Notice that this convolution is not commutative, for instance see terms such as $C_1\otimes P_1\otimes P_1$ in Eq.~(\ref{coef-zeta}), because the kernels are not symmetric in flavor space. The expression in Eq.~(\ref{small-b:position}) is independent of $\mu_{\text{OPE}}$ as the dependence on this scale cancels between PDF evolution and the coefficient function, rendering the lhs of Eq.~(\ref{small-b:position}) scale invariant. For the optimal TMD,  the perturbative coefficient function takes the following form:
\begin{eqnarray}\label{coef-zeta}
\nn \widetilde{C}&=& \mathbf{1}+\alpha_s \(-P_1 \mathbf{L}_{\mathcal{O}}+C_1\)
+\alpha_s^2 \[\frac{P_1\otimes P_1-\beta_0 P_1}{2}\mathbf{L}_{\mathcal{O}}^2
- \(P_2+C_1\otimes P_1-\beta_0 C_1\) \mathbf{L}_{\mathcal{O}}+C_2\]
\\ &&
+\alpha_s^3\Big[
-\(P_1\otimes P_1\otimes P_1-3\beta_0 P_1\otimes P_1+2\beta_0^2P_1\)\frac{\mathbf{L}_{\mathcal{O}}^3}{6}
\\\nn &&+\(P_1\otimes P_2+P_2\otimes P_1+C_1\otimes P_1\otimes P_1-3\beta_0 C_1\otimes P_1-2\beta_0 P_2-\beta_1 P_1+2\beta_0^2 C_1\)\frac{\mathbf{L}_{\mathcal{O}}^2}{2}
\\\nn &&-\(P_3+C_1\otimes P_2+C_2\otimes P_1-2\beta_0C_2-\beta_1 C_1\)\mathbf{L}_{\mathcal{O}}+C_3\Big]+\mathcal{O}(\alpha_s^4), 
\end{eqnarray}
where the leading term $\mathbf{1} \equiv \delta_{ff'}\delta(1-y)$ represents the unity convolution, $C_n$ are the finite parts of the coefficient function free from any logarithmic dependence, and $P_n$ stand for the perturbative coefficients of the DGLAP kernel, as given by
\begin{eqnarray}\label{def:DGLAP}
\mu^2 \frac{d}{d \mu^2} f(x,\mu)=\(\sum_{n=1}^\infty \alpha_s^n(\mu)P_n(x)\)\otimes f(x,\mu)=P\otimes f(x,\mu).
\end{eqnarray}
Furthermore, $\beta_n$ represent the coefficients of the QCD beta function, defined by
\begin{eqnarray}
\mu^2 \frac{d}{d\mu^2}\alpha_s(\mu)=-\sum_{n=0}^\infty \beta_n \alpha_s^{n+2}(\mu).
\end{eqnarray}

Expressions for the coefficient functions for the unpolarized TMD can be found in Refs.~\cite{Echevarria:2016scs, Luo:2019szz, Ebert:2020yqt, Luo:2020epw} up to $\alpha_s^3$ order, for helicity TMD in Refs.~\cite{Bacchetta:2013pqa, Gutierrez-Reyes:2017glx} up to $\alpha_s^1$, and for transversity TMD in Ref.~\cite{Gutierrez-Reyes:2018iod} up to $\alpha_s^2$. The expressions presented in these references apply to general scales $(\mu,\zeta)$, and in the Appendix we provide the rules to render those into $\zeta$ prescription. It is essential to emphasize that all computations are in the  $\overline{\text{MS}}$ scheme, and consequently, PDFs in Eq.~(\ref{small-b:position}) are defined in the $\overline{\text{MS}}$ scheme.

The relationship between the small-$b$ and large-$\mu$ asymptotic expansions with logarithmic terms is explored in detail in Refs.~\cite{WONG1977271, MacKinnon:1972}.  Generally, power-suppressed terms in $b$ contribute to power-suppressed terms in $\mu$, and the logarithmic singularities at $b\to0$ turn into logarithmic singularities at $\mu\to\infty$. For the leading power term, the asymptotic behavior for $b\to0$ and $\mu\to\infty$ are connected by simple replacement rules $\mathbf{L}_{\mathcal{O}}\to -\pmb{{\ell}}$, $\mathbf{L}_{\mathcal{O}}^2\to \pmb{{\ell}}^2$, $\mathbf{L}_{\mathcal{O}}^3\to -\pmb{{\ell}}^3-4\zeta_3$, etc., see Ref.~\cite{WONG1977271} (here $\pmb{{\ell}}=\ln(\mu^2/\mu^2_{OPE})$). The OPE in Eq.~(\ref{small-b:position}) is independent of $\mu_{\text{OPE}}$, and therefore, it is conventional to set $\mu_{\text{OPE}}=\mu$ (and hence $\pmb{{\ell}}=0$). Thus, the asymptotic form of Eq.~(\ref{G0=f}) reads as
\begin{eqnarray}\label{G0[f]=expansion}
&&\mathcal{G}_0[F](x,\mu)=\bigg\{\mathbf{1}+\alpha_s C_1
+\alpha_s^2 C_2
\\\nn && \qquad 
+\alpha_s^3\Big[
\frac{2\zeta_3}{3}\(P_1\otimes P_1\otimes P_1-3\beta_0 P_1\otimes P_1+2\beta_0^2P_1\)+C_3\Big]+\mathcal{O}(\alpha_s^4)\bigg\}\otimes f(x,\mu)+\mathcal{O}(\mu^{-2}),
\end{eqnarray}
where $\alpha_s$ is evaluated at $\mu$. Upon differentiating this expression, we obtain
\begin{eqnarray}
\mu^2 \frac{d}{d\mu^2}f^{(\text{TMD})}(x,\mu)=P'\otimes f^{(\text{TMD})}(x,\mu).
\end{eqnarray}
Here, the evolution kernel $P'$ deviates from the $\overline{\text{MS}}$ DGLAP kernel $P$, defined in Eq.~(\ref{def:DGLAP}), starting at order $\alpha_s^2$:
\begin{eqnarray}\label{DGLAP-difference}
P'-P=-\alpha_s^2 \beta_0 C_1-\alpha_s^3\(2\beta_0 C_2-\beta_0 C_1\otimes C_1+\beta_1 C_1\)+\mathcal{O}(\alpha_s^4).
\end{eqnarray}
Thus, we have established that the function $f^{(\text{TMD})}$ is a collinear PDF that obeys the DGLAP equation but is computed in a scheme different from $\overline{\text{MS}}$. As we stated in the previous subsection, we refer to this scheme as the TMD scheme. The same result holds for helicity and transversity zeroth TMM, employing the pertinent DGLAP kernels $P$ and coefficient functions $C$.

The transition from the TMD scheme to the $\overline{\text{MS}}$ scheme is accomplished through multiplication by a finite renormalization matrix $Z$:
\begin{eqnarray}\label{finite-renormalization}
f^{(\MS)}_f(x,\mu)&=&\sum_{f'}\int_x^1 \frac{dy}{y}Z^{\MS/\text{TMD}}_{f\ot f'}(y,\mu)f_{f'}^{(\text{TMD})}\(\frac{x}{y},\mu\),
\end{eqnarray}
where the PDF on the left-hand side is the collinear PDF in the $\overline{\text{MS}}$ scheme (explicitly indicated by the superscript $\overline{\text{MS}}$). Expressions for the finite renormalization constant $Z^{\MS/\text{TMD}}$ can be obtained by eliminating terms in Eq.~(\ref{DGLAP-difference}). The result reads as 
\begin{eqnarray}\label{Z:TMD->MS}
&&Z^{\MS/\text{TMD}}=\mathbf{1}-\alpha_s C_1-\alpha_s^2 \(C_2-C_1\otimes C_1\)
\\\nn &&\qquad
-\alpha_s^3\[C_3+C_1\otimes C_1\otimes C_1-C_1\otimes C_2-C_2\otimes C_1
+\frac{2\zeta_3}{3}
P_1\otimes (P_1-\beta_0\cdot \mathbf{1})\otimes (P_1-2\beta_0\cdot \mathbf{1})\]+\mathcal{O}(\alpha_s^4).
\end{eqnarray}
The inverse factor can be expressed as
\begin{eqnarray}\label{Z:MS->TMD}
&&\(Z^{\MS/\text{TMD}}\)^{-1}=\mathbf{1}+\alpha_s C_1+\alpha_s^2 C_2+\alpha_s^3\[C_3+\frac{2\zeta_3}{3}
P_1\otimes (P_1-\beta_0\cdot \mathbf{1})\otimes (P_1-2\beta_0\cdot \mathbf{1})\]+\mathcal{O}(\alpha_s^4).
\end{eqnarray}

The present state-of-the-art DGLAP equation precision is next-to-next-to-leading order (NNLO), making the terms of order $\alpha_s^2$ in Eqs.~(\ref{Z:TMD->MS}) and (\ref{Z:MS->TMD}) sufficient for that level of accuracy.   For the sake of completeness, we present our results up to $\alpha_s^3$.

Hence, using Eqs.~(\ref{G0=f}) and (\ref{Z:TMD->MS}), one can determine collinear PDFs based on known values of TMDs. We emphasize that the scheme-transformation factor $Z$ is a matrix in flavor space. Consequently, to reconstruct a collinear PDF with a precision beyond NLO, both quark and gluon TMDs are required.

\subsection{TMDs at general scales}

If one employs a set of arbitrary TMD scales $(\mu_{\text{TMD}},\zeta)$, then, in general, it is not possible to reconstruct a collinear PDF. The reason is the double-logarithm form of the coefficient function, which has an auxiliary arbitrary scale $\mu_{\text{OPE}}$ and which produces an extra logarithmic contribution into the evolution of $\mathcal{G}_0[F]$ already at leading order (LO). These terms cannot be removed by any finite renormalization. However, in the special case of
\begin{eqnarray}\label{mu=...=zeta}
\mu=\mu_{\text{OPE}}=\mu_{\text{TMD}}=\sqrt{\zeta},  
\end{eqnarray}
the collinear PDF is obtained in a \textit{TMD2 scheme}, which can be converted to the $\overline{\text{MS}}$ scheme through a finite renormalization constant, different from the one obtained in  Eq.~(\ref{Z:TMD->MS}).

The most general coefficient function of the small-$b$ OPE for the TMD $F(x,b;\mu_{\text{TMD}},\zeta)$ involves three scales\footnote{Notice that $\mu_{\text{OPE}}$ dependence cancels with that in the evolution of the collinear PDFs in Eq.(\ref{small-b:position}).}: $\mu_{\text{OPE}}$, $\mu_{\text{TMD}}$, and $\zeta$. The expression for it has the following structure
\begin{eqnarray}\label{C-general}
&&C(\mathbf{L}_{\mathcal{O}},\mathbf{L}_T,\pmb{{\ell}}_T)=
\mathbf{1}+\alpha_s\left\{\(-\frac{\Gamma_0}{4}\mathbf{L}_T^2+\frac{\Gamma_0}{2}\mathbf{L}_T\pmb{{\ell}}_T-\frac{\gamma_1}{2}\mathbf{L}_T\)\cdot \mathbf{1}-P_1 \mathbf{L}_{\mathcal{O}}+\overline{C}_1\right\} 
\\\nn &&
\qquad
+\alpha_s^2\bigg\{\left[\frac{\Gamma_0^2}{32}\mathbf{L}_T^2
(\mathbf{L}_T-2\pmb{{\ell}}_T)^2
+\frac{\Gamma_0\gamma_1}{8}\mathbf{L}_T^2(\mathbf{L}_T-2\pmb{{\ell}}_T)
+\frac{\Gamma_0\beta_0}{12}\mathbf{L}_T
\(\mathbf{L}_T^2-3 \mathbf{L}_T\pmb{{\ell}}_T
-3\mathbf{L}_T\mathbf{L}_{\mathcal{O}}+6 \mathbf{L}_{\mathcal{O}}\pmb{{\ell}}_T\)\right.
\\\nn &&
\qquad\left.
+\mathbf{L}_T^2 \frac{\gamma_1^2+2\gamma_1\beta_0}{8}
-\mathbf{L}_T(\mathbf{L}_T-2\pmb{{\ell}}_T) \frac{\Gamma_1+\Gamma_0\overline{C}_1}{4}
-\mathbf{L}_T\frac{\gamma_2+2d^{(2,0)}+\gamma_1\overline{C}_1}{2}+\pmb{{\ell}}_T d^{(2,0)}\right]\cdot \mathbf{1}
\\\nn &&
\qquad
+\frac{\Gamma_0}{4}\mathbf{L}_T\mathbf{L}_{\mathcal{O}}\(\mathbf{L}_T-2\pmb{{\ell}}_T\)P_1
+\mathbf{L}_T\mathbf{L}_{\mathcal{O}} \frac{\gamma_1}{2}(P_1-\beta_0\cdot \mathbf{1})
+
\frac{1}{2}\mathbf{L}_{\mathcal{O}}^2 P_1\otimes (P_1-\beta_0\cdot \mathbf{1})
\\\nn &&
\qquad
-\mathbf{L}_{\mathcal{O}} (P_2+\overline{C}_1\otimes(P_1-\beta_0\cdot \mathbf{1}))+\overline{C}_2
\bigg\}+\mathcal{O}(\alpha_s^3),
\end{eqnarray}
where $\alpha_s$ is at the scale $\mu_{\text{OPE}}$, and
\begin{eqnarray}
\mathbf{L}_{\mathcal{O}}=\ln\(\frac{\mu_{\text{OPE}}^2 b^2}{4e^{-2\gamma_E}}\),
\qquad
\mathbf{L}_T=\ln\(\frac{\mu_{\text{TMD}}^2 b^2}{4e^{-2\gamma_E}}\),
\qquad
\pmb{{\ell}}_T=\ln\(\frac{\mu_{\text{TMD}}^2}{\zeta}\).    
\end{eqnarray}
Note that the finite parts of the coefficient function $\overline{C}_i$ are different from $C_i$ for the optimal TMD. The relation between them is provided in Eq.~(\ref{eq:ccbar}) in the Appendix.

The evaluation of $\mathcal{G}_0$ through the OPE effectively replaces $\ln b$ by $\ln \mu$, where $\mu$ is the integral cutoff in the operation $\mathcal{G}_0$. Therefore, the resulting function depends on the scales $\mu$, $\mu_{\text{TMD}}$, and $\zeta$. Importantly, the dependence on $\mu$ does not reproduce the DGLAP equation.  Setting $\mu_{\text{OPE}} = \mu$ we obtain
\begin{eqnarray}
\nn& &\mu^2 \frac{d}{d\mu^2}\mathcal{G}_0[F](x,\mu,\mu_{\text{TMD}},\zeta)=\bigg\{
\Big(
P(\mu)+\frac{\gamma_V(\mu)-\Gamma_{\text{cusp}}(\mu)\pmb{{\ell}}_\mu}{2}
\cdot \mathbf{1}
\Big)+\alpha_s^2\Big[\Big(\frac{\Gamma_0^2}{8} \pmb{{\ell}}_\mu 
(\pmb{{\ell}}_\mu^2-\pmb{{\ell}}_T^2)\\
\nn&&
\qquad
-\frac{\Gamma_0\gamma_1}{8}(\pmb{{\ell}}_\mu-\pmb{{\ell}}_T) (3\pmb{{\ell}}_\mu+\pmb{{\ell}}_T)
+\frac{\gamma_1^2}{4}(\pmb{{\ell}}_\mu-\pmb{{\ell}}_T)+d^{(2,0)}+\frac{\Gamma_0^2}{2}\zeta_3\Big)\cdot \mathbf{1}+\frac{1}{4}\Gamma_0(\pmb{{\ell}}_T^2-\pmb{{\ell}}_\mu^2)P_1
\\
&&\qquad
+\gamma_1(\pmb{{\ell}}_\mu-\pmb{{\ell}}_T)P_1-\frac{\Gamma_0}{2}\pmb{{\ell}}_\mu \overline{C}_1+\overline{C}_1\otimes P_1+\(\frac{\gamma_1}{2}-\beta_0\)\overline{C}_1
\Big]+\mathcal{O}(\alpha_s^3)\bigg\}\otimes f(x,\mu),
\end{eqnarray}
where $\alpha_s\equiv \alpha_s(\mu)$, and we also define
\begin{eqnarray}
\pmb{{\ell}}_\mu=\ln\(\frac{\mu^2}{\zeta}\).   
\end{eqnarray}
This equation does not exhibit the DGLAP structure and cannot be reduced to it through any finite factor, due to the $\pmb{{\ell}}_\mu$ term which is present at LO. Additionally, this function depends on $\mu_{\text{TMD}}$ and $\zeta$, as encoded in Eqs.~(\ref{evol:UV}) and (\ref{evol:rapidity}).

On the other hand, these issues can be avoided by setting scales as specified in Eq.~(\ref{mu=...=zeta}). This choice of scales significantly simplifies the logarithmic structure of the coefficient function, Eq.~(\ref{C-general}), and the evolution equation now becomes
\begin{eqnarray}
\mu^2 \frac{d}{d\mu^2}f^{(\text{TMD2})}(x,\mu)=\overline{P}\otimes f^{(\text{TMD2})}(x,\mu),
\end{eqnarray}
where we denote $f^{(\text{TMD2})}(x,\mu) \equiv \mathcal{G}_0[F](x,\mu,\mu,\mu^2)$. The evolution kernel $\overline{P}$ deviates from the DGLAP kernel at order $\alpha_s^2$, akin to the difference highlighted in Eq.~(\ref{DGLAP-difference}):
\begin{equation}
\begin{split}
    \overline{P}-P=&-\alpha_s^2\beta_0\overline{C}_1-\alpha_s^3\[2\beta_0 \overline{C}_2-\beta_0 \overline{C}_1\otimes \overline{C}_1+\beta_1 \overline{C}_1-2\zeta_3\Gamma_0\beta_0\(P_1+\Big(\frac{\gamma_1}{2}-\frac{2\beta_0}{3}\Big)\cdot \mathbf{1}\)\]+\mathcal{O}(\alpha_s^4)
\end{split} 
\end{equation}
Thus, the label ``(TMD2)" signifies that the collinear PDF is evaluated in another TMD scheme distinct from the $\overline{\text{MS}}$ scheme and the previously discussed TMD scheme labeled ``(TMD)." As before, one can construct the finite renormalization constant to pass from $f^{(\text{TMD2})}$ to the  $\overline{\text{MS}}$ scheme PDF. The  expression obtained for this constant is

\begin{eqnarray}\label{Z:TMD2->}
Z^{\MS/\text{TMD2}}=\mathbf{1}-\alpha_s \overline{C}_1-\alpha_s^2 \[\overline{C}_2-\overline{C}_1\otimes\overline{C}_1-\zeta_3\Gamma_0\(P_1+\Big(\frac{\gamma_1}{2}-\frac{2\beta_0}{3}\Big)\cdot \mathbf{1}\)\]+\mathcal{O}(\alpha_s^3),
\end{eqnarray}
and the inverse factor reads as

\begin{eqnarray}\label{Z:Ms->TMD2}
\(Z^{\MS/\text{TMD2}}\)^{-1}=\mathbf{1}+\alpha_s \overline{C}_1+\alpha_s^2 \[\overline{C}_2-\zeta_3\Gamma_0\(P_1+\Big(\frac{\gamma_1}{2}-\frac{2\beta_0}{3}\Big)\cdot \mathbf{1}\)\]+\mathcal{O}(\alpha_s^3).
\end{eqnarray}
This conclusion aligns with the findings of the authors in Ref. \cite{Ebert:2022cku}, where the agreement of $f^{(\text{TMD2})}(x,\mu)$ with the collinear PDF was demonstrated numerically.

Notice that also in this case our result stands for the unpolarized TMD, helicity TMD, and transversity TMD.

\subsection{Phenomenological examples}

\begin{figure}[t]
\centering
\includegraphics[width=0.94\textwidth]{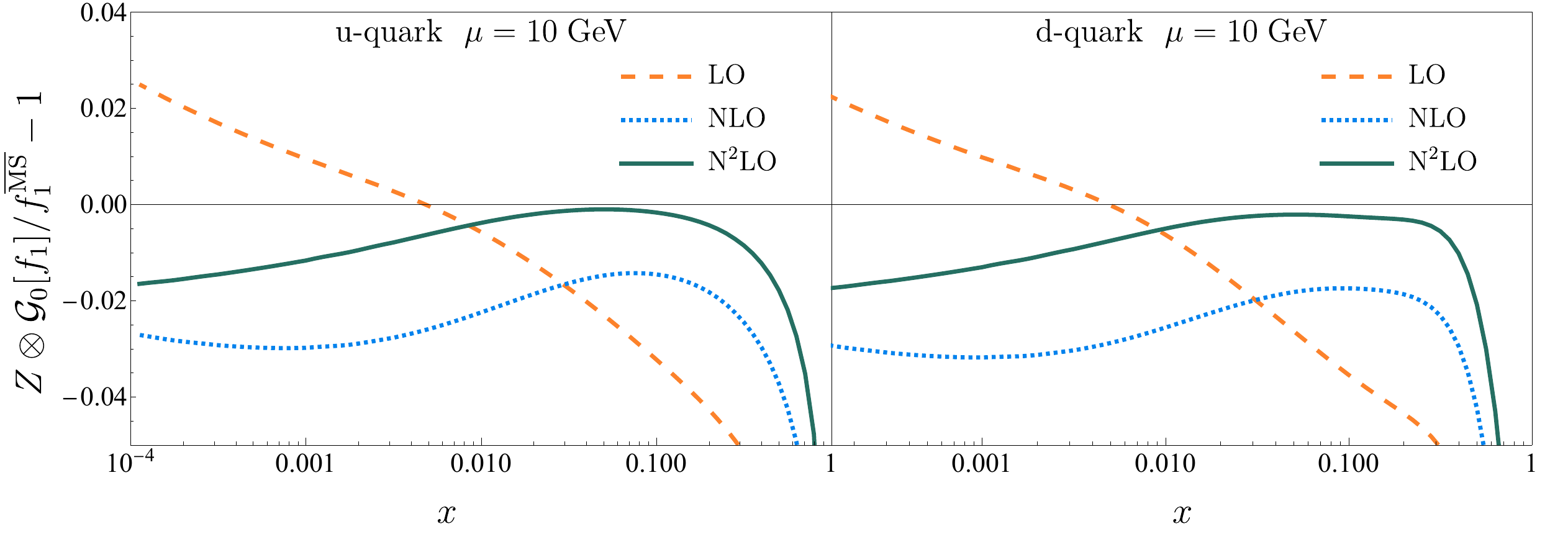}
\\
\includegraphics[width=0.94\textwidth]{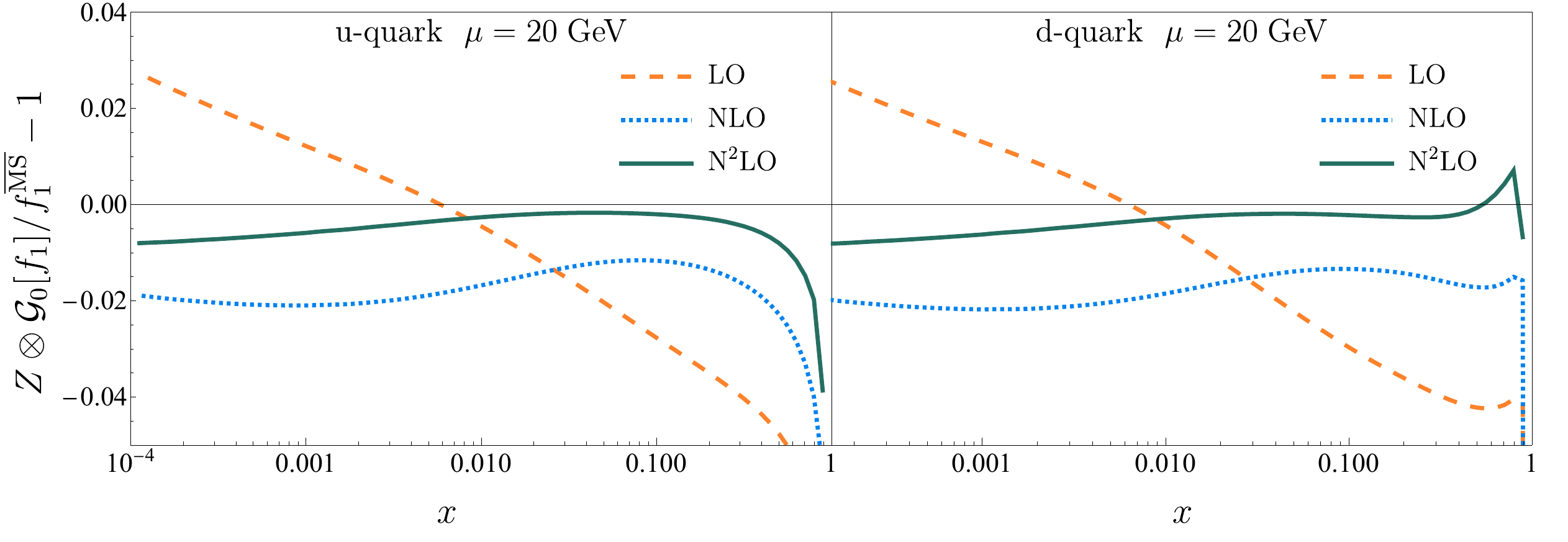}
\caption{Comparison of unpolarized PDF for $u$ and $d$ quarks determined from the unpolarized TMD (central values of ART23 extraction \cite{Moos:2023yfa}), as a function of $x$ at fixed $\mu=10$ (the upper row) and 20 GeV (the bottom row). The plots show the deviation from the $\overline{\text{MS}}$ value which was used in the fit of TMD (extraction MSHT20 \cite{Bailey:2020ooq}). Dashed orange lines, dotted blue lines, and solid green lines correspond to the LO, NLO, and NNLO  order of factor $Z^{\MS/\text{TMD}}$. 
\label{fig:TMDtoPDF_fixedMU}}
\end{figure}

\begin{figure}[ht]
\centering
\includegraphics[width=0.94\textwidth]{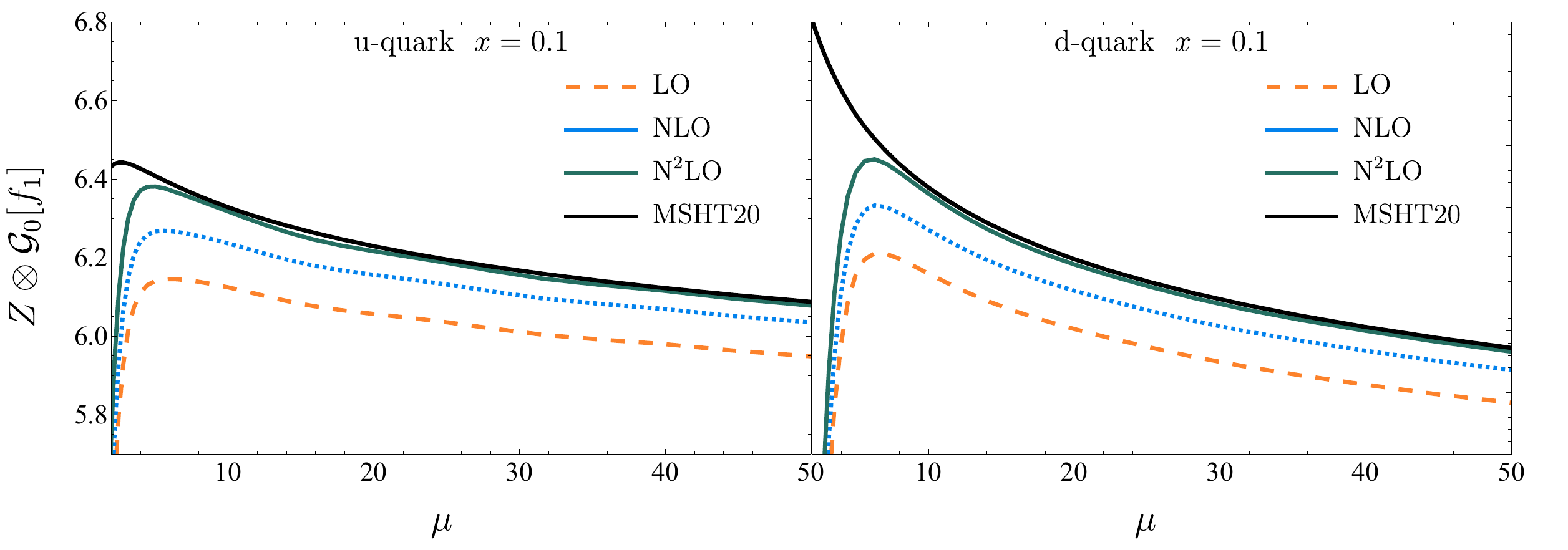}
\caption{Comparison of unpolarized PDF for u and d quarks determined from the unpolarized TMD (extraction ART23 \cite{Moos:2023yfa}), as a function of $\mu$ at fixed $x=0.1$. The plot shows the deviation from the $\overline{\text{MS}}$ value which was used in the fit of TMD (extraction MSHT20 \cite{Bailey:2020ooq}). Different lines correspond to different orders of correction factor $Z^{\MS/\text{TMD}}$.
\label{fig:TMDtoPDF_fixedX}}
\end{figure}

\begin{figure}[ht]
\centering
\includegraphics[width=0.94\textwidth]{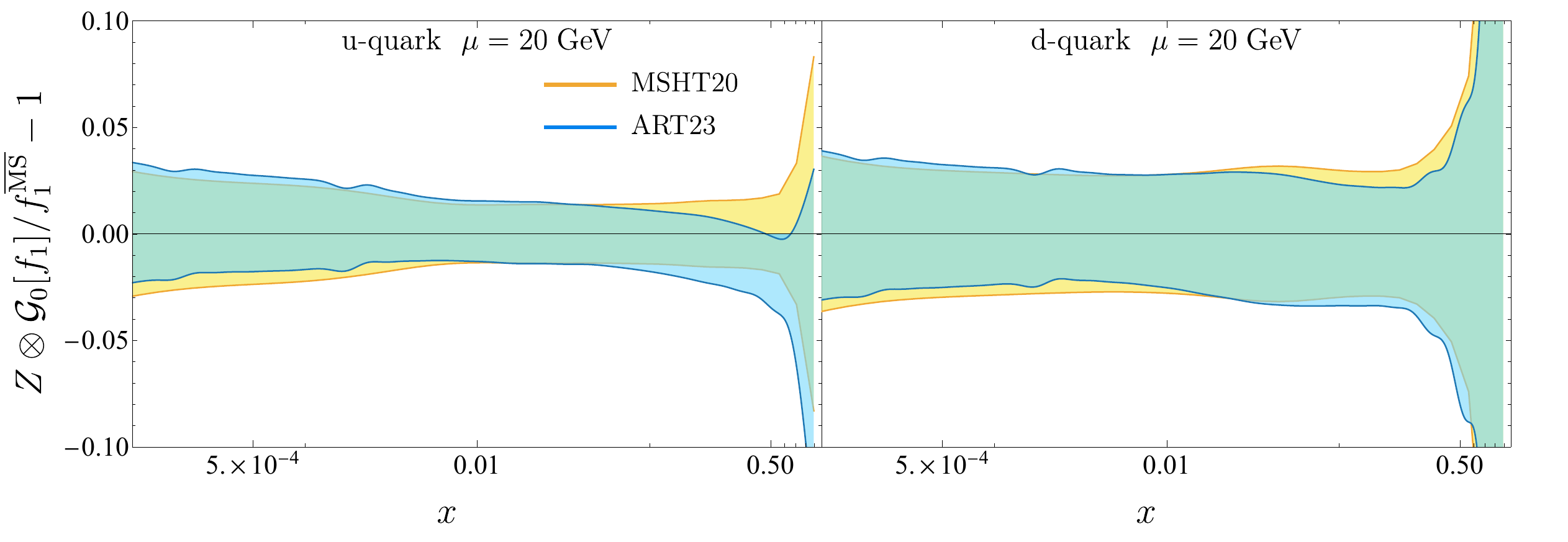}
\caption{Comparison of uncertainty bands for unpolarized PDF for u and d quarks as a function of $x$ at fixed $\mu=20$ GeV. The blue band is the uncertainty band determined from the uncertainty band of unpolarized TMD (extraction ART23 \cite{Moos:2023yfa}). The yellow band is the uncertainty band for unpolarized PDF (extraction MSHT20 \cite{Bailey:2020ooq}). Comparison is done with NNLO $Z^{\MS/\text{TMD}}$.
\label{fig:TMDtoPDF_unc}}
\end{figure}

To illustrate the application of our formulas and to verify their accuracy numerically, we consider a specific example of TMD extraction from experimental data, namely the ART23 determination of unpolarized TMD \cite{Moos:2023yfa}. ART23 analysis \cite{Moos:2023yfa} was performed at N4LL accuracy (with N$^3$LO matching to collinear PDF) via the global QCD fit of Drell-Yan and electroweak boson production data. The extraction has been done for optimal TMDs, and thus we use the formulas from~\sec{0th:optimal}.

In Figure~\ref{fig:TMDtoPDF_fixedMU} we plot $Z^{\MS/\text{TMD}}\otimes\mathcal{G}_0[f_1](x,\mu) / f^{\overline{\text{MS}}}(x,\mu) -1$  at two values of $\mu=10$ and 20 GeV, considering LO, NLO, and NNLO precision for $Z$. One can see that the agreement between the zeroth TMM and collinear PDFs improves at higher values of the scale $\mu$. Leading order expression already gives a very good agreement within 5\%.  The agreement between reconstructed and original PDFs is approximately 5\% at $\mu=10$ GeV (depending on $x$), and about 2\% at $\mu=20$ GeV. As higher-order corrections to $Z^{\MS/\text{TMD}}$ are applied, the precision  improves significantly\footnote{For the application of $Z^{\MS/\text{TMD}}$ we need the gluon TMD, which is not presently known. Instead, we used a pure OPE term with a constant nonperturbative function.}.
With $Z^{\MS/\text{TMD}}$ taken at NNLO, the agreement is of order of 2\%-5\% at $\mu=10$GeV, $\lesssim 1\%$ at $\mu=20$GeV, and at the subpercentage level for larger $\mu$. We have verified that the central line (without convolution with $Z^{\MS/\text{TMD}}$) agrees with the results presented in Ref.~\cite{Ebert:2022cku}. In the region of large $x$, the deviations become larger due to large $\ln(1-x)$ contributions. Potentially, the agreement can be improved using threshold resummation methods \cite{Korchemsky:1992xv,Idilbi:2006dg}.

Figure~\ref{fig:TMDtoPDF_fixedX} shows $Z^{\MS/\text{TMD}}\otimes
\mathcal{G}_0[f_1](x,\mu)$ at $x = 0.1$ as a function of $\mu$ and compares it to  MSHT20 \cite{Bailey:2020ooq}. This figure demonstrates that $f^{(\text{TMD})}$  reproduces very well the evolution of collinear PDF. One can see that the agreement starts from $\mu\sim 5$ GeV. For lower values of the scale $\mu$, the power corrections are substantial, and caution should be exercised in the application of our formulas.

Lastly, Fig.~\ref{fig:TMDtoPDF_unc} illustrates that the uncertainty band of TMD reproduces the uncertainty band of collinear PDF. This is a feature of the ART23 extraction, which incorporates PDF uncertainty into the TMD uncertainty band.
Fig.~\ref{fig:TMDtoPDF_unc} provides an important consistency test demonstrating that the input PDF is recovered completely with the correct uncertainty band. Possibly one can consider this feature in a broader context of proposed joined fits of TMDs and PDFs. TMM discussed in this paper can be utilized as an additional consistency check for the output of such a fit for the mean values and for the uncertainty bands. Notice that in other extractions such as Refs.~\cite{Scimemi:2019cmh,Bacchetta:2019sam},  where the central replica of PDFs was used in the OPE, we expect the uncertainty band of TMM to become very small.

In the case of helicity and transversity, the same relations hold, however at present it is not feasible to present a phenomenological study of these relations. There are no extractions of helicity TMDs that can be compared to the collinear helicity PDF which is currently extracted~\cite{Bertone:2024taw} at NNLO accuracy. In the case of transversity PDF, the current extraction has substantial uncertainties (see for instance Ref.~\cite{Cocuzza:2023vqs} and references therein), and therefore the usage of our relations would not be meaningful for their comparison.

\section{The first Transverse Momentum Moment}
\label{sec:1st}
The first TMMs read as
\begin{eqnarray}
\nn
\mathcal{M}_{\mu}^{[\gamma^+]}(x,\mu) = \int^\mu d^2 {\bm k}_T {\bm k}_{T\mu} F^{[\gamma^+]}(x,k_T)&=& - \int^\mu d^2 {\bm k}_T {\bm k}_{T{\mu}} \epsilon^{\rho\nu}_T \frac{k_{T\rho} s_{T\nu}}{M} f_{1T}^\perp(x,k_T),
\\\label{def:first-moment}
\mathcal{M}_\mu^{[\gamma^+\gamma_5]}(x,\mu) = \int^\mu d^2 {\bm k}_T {\bm k}_{T\mu} F^{[\gamma^+\gamma^5]}(x,k_T)&=& - \int^\mu d^2 {\bm k}_T {\bm k}_{T\mu} \frac{(k_T \cdot s_T)}{M} g^\perp_{1T}(x,k_T),
\\\nn
\mathcal{M}_\mu^{[i\sigma^{\alpha+}\gamma^5]}(x,\mu) = \int^\mu d^2 {\bm k}_T {\bm k}_{T\mu} F^{[i\sigma^{\alpha+}\gamma^5]}(x,k_T)&=& \int^\mu d^2 {\bm k}_T {\bm k}_{T\mu}\frac{\lambda k_T^\alpha}{M} h_{1L}^\perp(x,k_T) -\int^\mu d^2 {\bm k}_T {\bm k}_{T\mu} \frac{\epsilon_T^{\alpha\rho}k_{T\rho}}{M}h_1^\perp(x,k_T).
\end{eqnarray}

The first TMMs are expressed via the transformation $\mathcal{G}_1\equiv \mathcal{G}_{1,1}$, Eqs.~(\ref{def:Gn}), and they are given by
\begin{eqnarray}\label{M1:+}
\mathcal{M}^{[\gamma^+]}_\mu(x,\mu)&=&-\epsilon_{T,\mu \nu}s^{\nu}_T M\mathcal{G}_1[f_{1T}^\perp](x,\mu),
\\\label{M1:5+}
\mathcal{M}^{[\gamma^+\gamma^5]}_\mu(x,\mu)&=&-s_{T\mu}M\mathcal{G}_1[g_{1T}^\perp](x,\mu),
\\\label{M1:t+}
\mathcal{M}^{[i\sigma^{\alpha+}\gamma^5]}_\mu(x,\mu)&=&-\lambda g_{T, \mu\alpha} M\mathcal{G}_1[h_{1L}^\perp](x,\mu)-\epsilon_{T,\mu\alpha}M\mathcal{G}_1[h_1^\perp](x,\mu).
\end{eqnarray}
These quantities can be interpreted as the average displacement of the transverse momentum of a parton within a polarized hadron \cite{Boer:1997nt,Boer:1997bw,Burkardt:2003yg,Burkardt:2004ur,Meissner:2007rx,Gamberg:2017jha}. Their nonzero values are a consequence of the presence of the spin, whether it be the spin of the proton or the parton. Notice that all functions in $b$ space have index $(1)$, and the operation $\mathcal{G}_{n,m}$ (\ref{def:Gn}) is performed with $n=m=1$. This corresponds to the first moment of TMDs in the momentum space from Eq.~(\ref{eq:nth-moment}).

The small-$b$ expansion structure for these TMDs differs from what was considered in ~\sec{0th} for $f_1$, $g_1$, and $h_1$. Schematically this OPE can be expressed as
\begin{eqnarray}\label{small-$b$:position:tw3}
\widetilde{F}(x,b)=\sum_t  [\widetilde{C}_{t}(\mathbf{L}_{\mathcal{O}})\otimes t](x)+\mathcal{O}(b^2),
\end{eqnarray}
where $t$ represents collinear distributions of twist two, twist three, or twist four, and $\otimes$ denotes an integral convolution in momentum fractions. Two notable differences between Eq.~(\ref{small-b:position}) and Eq.~(\ref{small-$b$:position:tw3}) are noteworthy. Firstly, twist-three and twist-four distributions $t$ depend on two or more collinear momentum fractions.  Therefore, the convolution with the coefficient function projects several variables onto a single variable $x$ in the lhs of Eq.~(\ref{small-$b$:position:tw3}). For example, the leading term for the small-$b$ expansion of the Sivers function is a projection~\cite{Ji:2006ub, Ji:2006vf} of twist-three distributions $T(x_1,x_2,x_3)$ onto the Qiu-Sterman function 
\begin{eqnarray}\label{QS}
\widetilde{f}_{1T}^{\perp (1)}(x,b)=\pm\pi T(-x,0,x)+\mathcal{O}(\alpha_s,b^2).
\end{eqnarray}
 where $\pm$ corresponds to either semi-inclusive deep inelastic scattering (SIDIS), $``-"$, or Drell-Yan, $``+"$. The corresponding coefficient function at the zeroth order in $\alpha_s$ is $\delta(x_2)\delta(x+x_1)\delta(x-x_3)$. The second difference is that the formula (\ref{small-$b$:position:tw3}) relates one TMD to several collinear distributions. For instance, the Sivers TMD $f_{1T}^\perp$ is related to the twist-three distribution $T(x_1,x_2,x_3)$, and (starting from NLO) to the twist-three distribution $\Delta T(x_1,x_2,x_3)$ \cite{Scimemi:2019gge}. The worm-gear-T TMD $g_{1T}^\perp$ is related to the twist-two helicity distributions $\Delta q$ and to the twist-three distributions $T$ and $\Delta T$ \cite{Rein:2022odl}, and so forth. Explicitly \cite{Scimemi:2018mmi, Rein:2022odl},
\begin{eqnarray}
\mathcal{G}_1[f_{1T}^\perp](x,\mu) &=&  \pm\frac{\pi}{2} T^{(\text{TMD})}(-x,0,x; \mu)+\mathcal{O}(\mu^{-2}), \nn \\\nn
\mathcal{G}_1[g_{1T}^\perp](x,\mu) &=&  \frac{x}{2}\int_x^1 \frac{dy}{y} \Delta q^{(\text{TMD})}(y,\mu)
+x \int_{-1}^1 dy_1dy_2dy_3 \delta(y_1+y_2+y_3)\int_0^1 d\alpha \delta(x-\alpha y_3)\Big[
\\ &&
\qquad\qquad
\frac{\Delta T^{(\text{TMD})}(y_{123};\mu)}{y_2^2}
+
\frac{T^{(\text{TMD})}(y_{123};\mu)-\Delta T^{(\text{TMD})}(y_{123};\mu)}{2y_2y_3}
\Big]
+\mathcal{O}(\mu^{-2}), \\\nn
\mathcal{G}_1[h_{1L}^\perp](x,\mu) &=&  -\frac{x^2}{2}\int_x^1 \frac{dy}{y} \delta q^{(\text{TMD})}(y,\mu)
\\\nn &&
-x \int_{-1}^1 dy_1dy_2dy_3 \delta(y_1+y_2+y_3)\int_0^1 d\alpha \alpha \delta(x-\alpha y_3)H^{(\text{TMD})}(y_{123};\mu)
\frac{y_3-y_2}{y_2^2y_3}
+\mathcal{O}(\mu^{-2}), \\\nn
\mathcal{G}_1[h_{1}^\perp](x,\mu) &=&  \mp\frac{\pi}{2} E^{(\text{TMD})}(-x,0,x; \mu)+\mathcal{O}(\mu^{-2}),
\end{eqnarray}
where the shorthand $y_{123} \equiv y_1,y_2,y_3$, functions $T$, $\Delta T$, $H$ and $E$ are twist-3 collinear PDFs whose explicit definitions can be found in Refs.~\cite{Braun:2021aon, Braun:2021gvv, Rein:2022odl}. The signs of the first and the last equation should be understood as Drell-Yan, the upper signs, and SIDIS, the lower sign.

These features are also characteristics of the collinear matrix elements $\mathbb{M}_\mu^{[\Gamma]}$ from Eq.~(\ref{def:Mn-operator}) which lack definite twist and consequently represent a mixture of different contributions.

The perturbative structure of coefficient $\widetilde{C}_t$ in the $\zeta$ prescription follows a form similar to Eq.~(\ref{coef-zeta}).  It can be written as
\begin{eqnarray}\label{Ctr}
\widetilde{C}_t=R_t\otimes\(\mathbf{1}+\alpha_s\(-P_{1t}\mathbf{L}_{\mathcal{O}}+C_{1t}\)+\mathcal{O}(\alpha_s^2)\),
\end{eqnarray}
where $R_t$ is the projection operator, and $P_t=\sum_n \alpha_s^n P_{nt}$. The projection operator $R_t$ projects the multivariable higher-twist distribution to the single variable $x$. For instance, in the case of the Sivers function the operation $R_t$ is $\pi\delta(x_2)\delta(x_1+x_2+x_3)\delta(x_3-x)$, which being integrated with twist-three distribution $T(x_1,x_2,x_3)$ results in Eq.~(\ref{QS}). The coefficients $C_{1t}$ are known for all TMDs except $h_{1T}^\perp$; see Refs.~\cite{Rein:2022odl, Scimemi:2019gge, Kang:2011mr, Sun:2013hua, Dai:2014ala}. The presence of the projection operator does not allow one to turn $\mathcal{M}_\mu$ to the $\overline{\text{MS}}$ scheme, as it would require a convolution with all variables $x_i$. Nevertheless, the difference between $\mathcal{M}_\mu$ in the TMD scheme and the $\overline{\text{MS}}$ scheme starts at NLO, alike for the zeroth TMM.  To demonstrate this, one can differentiate $\mathcal{M}_\mu$ with respect to the scale and from Eq.~(\ref{Ctr}) one obtains
\begin{eqnarray}\label{evol:TMD-vs-MS}
\mu^2 \frac{d}{d\mu^2}\mathcal{G}_1[F](x,\mu)=R_t\otimes P'_t\otimes t+\mathcal{O}(\alpha_s^2),
\end{eqnarray} 
where $F$ represents any of the TMDs appearing in Eqs.~(\ref{M1:+})-(\ref{M1:t+}). The right-hand side of Eq.~(\ref{evol:TMD-vs-MS}) is the LO evolution of $\mathcal{M}_\mu$ in the $\overline{\text{MS}}$ scheme. At NLO, there is a term containing the one-loop finite part $C_{1t}$, which deviates from the expression in the $\overline{\text{MS}}$ scheme [see Eq. (\ref{DGLAP-difference})], and so we have $P'_t-P_t = \mathcal{O}(\alpha_s^2)$. Note, that presently the twist-3 evolution kernel is known only at order $\alpha_s$. Thus, collinear distributions determined by TMMs are as precise as determinations by other methods.

There is one ``accidental'' exception from this general rule, namely, the Boer-Mulders function $h_1^\perp$ that has $C_{1t}$ that commutes with the projector $C_{1t}=-\zeta_2 C_F\mathbf{1}$ \cite{Rein:2022odl}. Also the Boer-Mulder function is chiral odd, and does not mix with twist-3 gluon distributions. Therefore, in this case, one can find the NLO part of the transformation factor between TMD and  $\MS$ schemes
\begin{eqnarray}
\(1+a_s(\mu)C_F\frac{\pi^2}{6}\)\mathcal{G}_1[h_1^\perp]=\mp \frac{\pi}{2} E(-x,0,x)(1+\mathcal{O}(a_s^2)),
\end{eqnarray}
where $E$ is the chiral-odd distribution of twist 3, see definition in Refs.~\cite{Braun:2021gvv, Rein:2022odl}. However, one cannot expect that such construction would be possible beyond NNLO.

Thus, the first TMM $\mathcal{M}_\mu$ is related to $\mathbb{M}_\mu$ computed in the  scheme.  Since the matrix element $\mathbb{M}_\mu$ is not an autonomous function (in the sense that it mixes with other functions by the QCD evolution), one cannot transform  $\mathcal{M}_\mu$ to the  $\overline{\text{MS}}$ scheme without using extra information. Certainly, the connections between the twist-3 and TMD functions present intriguing avenues for further theoretical and phenomenological investigations~\cite{Gamberg:2017jha,Qiu:2020oqr}.

\subsection{General scales}

At general scales, the OPE can be formally represented as
\begin{eqnarray}
\widetilde{F}(x,b;\mu_{\text{OPE}},\mu_{\text{TMD}},\zeta)&=&\sum_t[C_t(\mathbf{L}_{\mathcal{O}},\mathbf{L}_T,\pmb{{\ell}}_{T}) \otimes t](x)+\mathcal{O}(b^2),
\end{eqnarray}
where the summation is over collinear distributions of twist two, twist three, or twist four, and $\otimes$ denotes an integral convolution in momentum fractions. Analogously to the case discussed earlier, the coefficient $C_t$ contains the projection operator $R_t$, preventing the direct transformation of the first TMMs, $\mathcal{M}_\mu$, to the  $\overline{\text{MS}}$ scheme. Therefore, the persisting challenges in the transformation from the TMD scheme to the  $\overline{\text{MS}}$ scheme, previously observed for the optimal TMD case, continue to apply at general scales, even when considering the specific set of scales of Eq.~(\ref{mu=...=zeta}).

\subsection{Phenomenological example}

The Sivers function has attracted considerable attention in the literature~\cite{Boer:1997nt,Boer:1997bw,Burkardt:2003yg,Burkardt:2004ur,Meissner:2007rx,Gamberg:2017jha,Bury:2020vhj, Bury:2021sue}. This function is related~\cite{Ji:2006ub, Ji:2006vf} to the Qiu-Sterman function and provides a 3D snapshot of the transversely polarized nucleon.
One of the interesting features of the Sivers function is encoded in its first TMM $\mathcal{M}^{[\gamma^+]}_\mu$, as it can be interpreted as the average transverse momentum shift of partons with given $x$ due to the spin-orbital interactions~\cite{Burkardt:2003yg}.

\begin{figure}[t]
\centering
\includegraphics[width=0.99\textwidth]{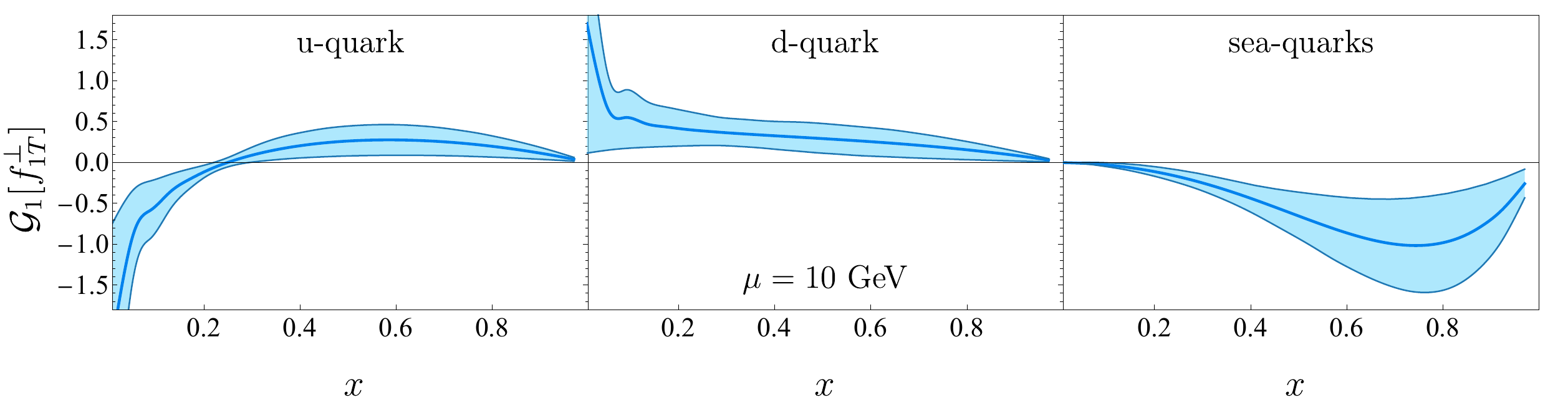}
\caption{The first TMM for the Sivers function (extraction \cite{Bury:2020vhj, Bury:2021sue}) for different flavors, computed at $\mu=10$ GeV.
\label{fig:sivers_g1}}
\end{figure}

We compute $\mathcal{M}^{[\gamma^+]}_\mu$ using recent extractions of the Sivers function at N$^3$LO~\cite{Bury:2020vhj, Bury:2021sue}. In these extractions, $u$, $d$, and $sea$ quark Sivers functions were determined. Figure~\ref{fig:sivers_g1} illustrates $\mathcal{M}_\mu^{[\gamma^+]}$ for $u$, $d$, and sea quarks based on Refs.~\cite{Bury:2020vhj, Bury:2021sue}.

If one integrates over $x$, the result can be interpreted as the mean transverse momentum shift of a parton in a transversely polarized hadron. Let us denote \cite{Burkardt:2003yg}
\begin{eqnarray}
\langle \vec k_{T,\nu}^f \rangle(\mu)=\int_0^1 dx \mathcal{M}^{[\gamma^+]}_{\nu,f}(x,\mu),
\end{eqnarray}
where $f$ is the flavor index. As seen from Eq.~(\ref{M1:5+}), for a hadron polarized in the $\hat y$ direction, we have $\nu = 1$. For the extraction from \cite{Bury:2021sue}, we obtain the following values at $\mu=10$ GeV:
\begin{eqnarray}\label{eq:shifts}
\langle \vec k_{T,1}^u \rangle=-0.011_{-0.023}^{+0.011}\;\text{GeV},\qquad
\langle \vec k_{T,1}^d \rangle=0.17_{-0.17}^{+0.21}\;\text{GeV},\qquad
\langle \vec k_{T,1}^{sea} \rangle=-0.26_{-0.32}^{+0.26}\;\text{GeV},
\end{eqnarray}
where the uncertainties are computed according to Ref.~\cite{Bury:2021sue}. It is worth noting that the errors in these quantities are correlated. These numbers are consistent with lattice QCD results from Ref.~\cite{Musch:2011er} that give the Sivers shift for $u-d$  quarks in the range of $\langle \vec k_{T,1}^{u-d} \rangle=-0.3\dots -0.15$ GeV. The results in Eq.~(\ref{eq:shifts}) are comparable\footnote{Notice that the definition of the observables $\langle k_\perp^q \rangle$ reported in Ref.~\cite{Anselmino:2008sga} differs from our Eq.~(\ref{eq:shifts}) by a minus sign.} to those extracted in  Ref.~\cite{Anselmino:2008sga} in a TMD parton model analysis of the transverse single spin asymmetries for pion and kaon production in Semi-Inclusive Deep Inelastic Scattering.

The so-called Burkardt sum rule \cite{Burkardt:2003yg} conjectures that the sum over all $\langle \vec k_{T,1} \rangle$ is zero. This and other sum rules for TMDs were studied in Ref.~\cite{Lorce:2015lna}. In our notations, this can be expressed as
\begin{eqnarray}
\sum_{f=q,\bar q,g} \int_0^1 dx \mathcal{M}^{[\gamma^+]}_{\nu,f}(x,\mu) = \sum_{f=q,\bar q,g} \langle \vec k_{T,\nu}^f \rangle = 0.
\end{eqnarray}
While the Burkardt sum rule is not formally proven in QCD, it has been verified in model computations, as shown in previous studies \cite{Goeke:2006ef, Courtoy:2008dn}. If we define the following sum as
\begin{eqnarray}\label{budkardt}
\sum_{f=q,\bar q,g} \langle \vec k_{T,\nu}^f \rangle=\overline{\langle \vec k_{T,\nu} \rangle} \; ,
\end{eqnarray}
then, intriguingly, the Burkardt sum rule exhibits autonomous evolution at LO \cite{Zhou:2015lxa}
\begin{eqnarray}
\mu^2 \frac{d}{d\mu^2}\overline{\langle \vec k_{T,\nu} \rangle}=-\frac{\alpha_s}{2\pi}C_A \overline{\langle \vec k_{T,\nu} \rangle}+\mathcal{O}(\alpha_s^2).
\end{eqnarray}
Therefore, if $\overline{\langle \vec k_{T,\nu} \rangle}$ is zero at the initial scale, it continues to be zero at other scales. For the specific case of the extraction from Ref.~\cite{Bury:2021sue}, we find that, at $\mu=10$ GeV, the result for the sum over quark flavors is
\begin{eqnarray}
\sum_{f=q,\bar q}\langle \vec k^f_{T,1} \rangle=\overline{\langle \vec k_{T,1} \rangle}-
\langle \vec k_{T,1}^g \rangle=-0.14_{-0.31}^{+0.14} \text{GeV}.
\end{eqnarray}
This allows us to estimate the contribution of the gluon Sivers function as $\langle \vec k_{T,1}^g \rangle \simeq 0.14^{+0.31}_{-0.14}$ GeV, assuming the validity of the Burkardt sum rule, Eq.~(\ref{budkardt}). Therefore based on Ref.~\cite{Bury:2021sue}, one could anticipate potentially sizable gluon Sivers functions, comparable in magnitude to the Sivers function for the $d$ quark, even though the error band is large.

\section{The second Transverse Momentum Moment}
\label{sec:2nd}

Now, we proceed to derive expressions for the second TMMs:
\begin{eqnarray}
\nn
\mathcal{M}^{[\gamma^+]}_{\mu\nu}(x,\mu) = \int^\mu d^2 {\bm k}_T {\bm k}_{T\mu} {\bm k}_{T\nu} F^{[\gamma^+]}(x,k_T)&=& \int^\mu d^2 {\bm k}_T {\bm k}_{T\mu} {\bm k}_{T\nu}f_1(x,k_T),
\\\label{def:second-moment}
\mathcal{M}^{[\gamma^+\gamma_5]}_{\mu\nu}(x,\mu) = \int^\mu d^2 {\bm k}_T {\bm k}_{T\mu} {\bm k}_{T\nu} F^{[\gamma^+\gamma^5]}(x,k_T)&=& \lambda \int^\mu d^2 {\bm k}_T {\bm k}_{T\mu} {\bm k}_{T\nu} g_{1}(x,k_T),
\\\nn
\mathcal{M}^{[i\sigma^{\alpha+}\gamma^5]}_{\mu\nu}(x,\mu) = \int^\mu d^2 {\bm k}_T {\bm k}_{T\mu} {\bm k}_{T\nu} F^{[i\sigma^{\alpha+}\gamma^5]}(x,k_T)&=& s_T^\alpha \int^\mu d^2 {\bm k}_T {\bm k}_{T\mu} {\bm k}_{T\nu} h_{1}(x,k_T) 
\\\nn &&
- \int^\mu d^2 {\bm k}_T {\bm k}_{T\mu} {\bm k}_{T\nu} \frac{{\bm k}_T^2}{M^2}
\(\frac{g_T^{\alpha\rho}}{2}+\frac{k_T^\alpha k_T^\rho}{{\bm k}_T^2}\)s_{T\rho}h_{1T}^\perp(x,k_T).
\end{eqnarray}

By considering the operation defined in Eq.~(\ref{def:Gnm}) with $n\rightarrow n+1$ and $m\rightarrow n$, we obtain
\begin{eqnarray}\label{def:Gnplus1n}
\mathcal{G}_{n+1,n}[F](x,\mu)&=&\int^\mu d^2\vec k_T \left(\frac{\vec k_T^2}{2 M^{2}}\right)^{n+1} F(x,k_T)
\\\nn &=&\frac{1}{2M^2 n!}\int_0^\infty db\,\mu^3\left(\frac{\mu b}{2}\right)^n \frac{(n+1)J_{n+1}(\mu b)-J_{n+3}(\mu b)}{n+2}\widetilde{F}^{(n)}(x,b),
\end{eqnarray}

which, for the second moment, contributes solely with indices $(1,0)$, leading to

\begin{eqnarray}\label{M2:+}
\mathcal{M}^{[\gamma^+]}_{\mu\nu,\text{div}}(x,\mu)&=&-g_{T,\mu\nu} M^2 \mathcal{G}_{1,0}[f_{1}],
\\
\mathcal{M}^{[\gamma^+\gamma^5]}_{\mu\nu,\text{div}}(x,\mu)&=&-\lambda g_{T,\mu\nu} M^2\mathcal{G}_{1,0}[g_{1}],
\\\label{M2:t+}
\mathcal{M}^{[i\sigma^{\alpha+}\gamma^5]}_{\mu\nu,\text{div}}(x,\mu)&=&s_{T,\alpha}   g_{T,\mu\nu} M^2 \mathcal{G}_{1,0}[h_{1}]+\(g_{T,\mu\alpha}s_{T,\nu}+g_{T,\nu\alpha}s_{T,\mu}-g_{T,\mu\nu} s_{T,\alpha}\)\frac{M^2}{2}\mathcal{G}_{2}[h_{1T}^\perp].
\end{eqnarray}
Notice that the contributions of $f_1$, $g_1$, and $h_1$ involve the operation $\mathcal{G}_{1,0}$, while $h_{1T}^\perp$  contributes to the second TMM through $\mathcal{G}_{2}$. The subscript ``div'' in Eqs.~(\ref{M2:+})-(\ref{M2:t+}), indicates that these functions exhibit power divergences at large $\mu$, as discussed below. The very last term of Eq.~(\ref{M2:t+}) corresponds to the collinear counterpart of the pretzelocity. It incorporates twist-three and twist-four distributions \cite{Moos:2020wvd} and reproduces the $\overline{\text{MS}}$ expression up to NLO. Its $\ln\mu$ divergence is the same UV divergence as in the corresponding twist-4 operator, and doing the same derivations as in~\sec{0th}, we conclude that this TMM reproduces the collinear counterpart of the pretzelocity $h_{1T}^\perp$ function.

To establish a connection between Eqs.~(\ref{M2:+})-(\ref{M2:t+}) and the matrix elements in Eq.~(\ref{def:Mn-operator}), it is necessary to examine the small-$b$ OPE up to the first power of $b^2$. The general form of this expansion is given by
\begin{eqnarray}\label{OPE:withb4}
\widetilde{F}(x,b)&=&\widetilde{C}\otimes f+b^2 \sum_{k} \widetilde{C}^{(2)}_k\otimes f_k^{\text{tw 4}}+\mathcal{O}(b^4),
\end{eqnarray}
where the first term is presented in Eq.~(\ref{small-b:position}), and the second term represents the power correction, typically involving distributions of twist four and target mass corrections. The coefficients $\widetilde{C}^{(2)}$ are presently unknown (except for the target-mass corrections part at the tree order \cite{Moos:2020wvd}). Notably, at the tree order, the $b^2$ term in the OPE, Eq.~(\ref{OPE:withb4}), is equivalent to the matrix element $\mathbb{M}$ in Eq.~(\ref{def:Mn-bare}) (with appropriately contracted indices).  Thus, to relate $\mathcal{M}_{\mu\nu,\text{div}}$ and $\mathbb{M}_{\mu\nu}$, it is necessary to subtract the leading small-$b$ contribution, $\widetilde{C}\otimes f$.

The integrals of the type $\mathcal{G}_{n+1,n}$ behave as $\propto \mu^2$ at large $\mu$ reproducing the UV power divergence of the corresponding operator $\mathbb{M}_{\mu\nu}$, which is defined in Eq.(\ref{def:Mn-operator}). These power divergences are natural for physical renormalization schemes, such as the momentum subtraction scheme.  However, in the dimensional regularization and, consequently, in the  $\overline{\text{MS}}$ scheme, these power divergences do not appear directly.\footnote{These power divergences correspond in the  $\overline{\text{MS}}$ scheme to the so-called renormalons which appear due to the factorial growth of the coefficient functions. See the discussion in Ref.~\cite{Beneke:1998ui}, and an example of explicit computation in Ref.~\cite{Braun:2004bu}.} Therefore, to match with the $\overline{\text{MS}}$ scheme, one needs to subtract the leading $\propto \mu^2$ behavior from $\mathcal{M}_{\mu\nu}$. The resulting function $\mathcal{M}_{\mu\nu}$ is the matrix element $\mathbb{M}_{\mu\nu}$ computed in a minimal subtraction scheme (TMD scheme), which coincides with $\overline{\text{MS}}$ up to NLO. This relation is proven analogously to the approach used in the previous sections.

The asymptotic part of $\mathcal{G}_{n+1,n}$ can be computed analytically. We express the $\mathcal{G}_{n+1,n}$ transformation as
\begin{eqnarray}
\mathcal{G}_{n+1,n}[F](x,\mu)=\frac{\mu^2}{2 M^2} \text{AS}[\mathcal{G}_{n+1,n}[F]](x,\mu)+\overline{\mathcal{G}}_{n+1,n}[F](x,\mu),
\end{eqnarray}
where $\text{AS}[\mathcal{G}_{n+1,n}[F]]$ denotes the leading asymptotic term that can be derived from the first term in the OPE, Eq.~(\ref{OPE:withb4}). The term $\overline{\mathcal{G}}_{n+1,n}$ corresponds to $\mathbb{M}_{\mu\nu}$ in the TMD scheme, and it undergoes logarithmic evolution. Hence, the second moments in the $\overline{\text{MS}}$ scheme (up to NLO) are given by
\begin{eqnarray}
\mathcal{M}^{[\gamma^+]}_{\mu\nu}(x,\mu)&=&- g_{T,\mu\nu} M^2 \overline{\mathcal{G}}_{1,0}[f_{1}](x,\mu),
\\
\mathcal{M}^{[\gamma^+\gamma^5]}_{\mu\nu}(x,\mu)&=&-\lambda g_{T,\mu\nu} M^2\overline{\mathcal{G}}_{1,0}[g_{1}](x,\mu),
\\
\mathcal{M}^{[i\sigma^{\alpha+}\gamma^5]}_{\mu\nu}(x,\mu)&=&s_{T,\alpha}g_{T,\mu\nu} M^2 \overline{\mathcal{G}}_{1,0}[h_{1}](x,\mu)+\(
g_{T,\mu\alpha}s_{T,\nu}+g_{T,\nu\alpha}s_{T,\mu}-g_{T,\mu\nu} s_{T,\alpha}\)\frac{M^2}{2}\mathcal{G}_2[h_{1T}^\perp](x,\mu).
\end{eqnarray}
The asymptotic term $\text{AS}[\mathcal{G}_{1,0}[F]]$ for the cases $f_1$, $g_1$, and $h_1$ can be derived straightforwardly using Eq.~(\ref{coef-zeta}). Up to N$^3$LO, the expression reads as
\begin{eqnarray} \nn
\text{AS}[\mathcal{G}_{1,0}[F]](x,\mu)&=&\Big\{\alpha_sP_1+\alpha_s^2\[P_2+\(C_1-P_1\)\otimes\(P_1-\beta_0\mathbf{1}\)\]
\\\nn &&+\alpha_s^3\Big[P_3+\(C_1-P_1\)\otimes\(P_2-\beta_1\mathbf{1}\)
+\(C_2-P_2\)\otimes\(P_1-2\beta_0\mathbf{1}\)
\\\label{def:AS_X0} &&-\(C_1-P_1\)\otimes\(P_1-\beta_0\mathbf{1}\)\otimes\(P_1-2\beta_0\mathbf{1}\)\Big]
+\mathcal{O}\(\alpha_s^4\)\Big\}\otimes f(x,\mu),
\end{eqnarray}
where $f$ is the collinear PDF corresponding to $F$. This expression can be numerically computed using existing codes for TMD phenomenology \footnote{
Most codes for TMD phenomenology, see e.g. Ref.~\cite{artemide}, include the computation of the leading terms for the small-$b$ OPE in position space. In order to obtain Eq.~(\ref{def:AS_X0}), one should replace the logarithmic terms of the coefficient function in Eq.~(\ref{coef-zeta}) according to the rule
$$
\mathbf{L}_{\mathcal{O}}^0\to0,\qquad \mathbf{L}_{\mathcal{O}} \to -1,\qquad \mathbf{L}^2_{\mathcal{O}} \to -2,\qquad \mathbf{L}^3_{\mathcal{O}} \to -3!,\qquad \mathbf{L}_{\mathcal{O}}^4 \to -4!+16 \zeta_3,\qquad \text{etc.},
$$
where $\mathbf{L}_{\mathcal{O}}^0$ corresponds to the logarithmless part of the coefficient function.
}.

Therefore, armed with the knowledge of the optimal TMD and the corresponding collinear PDF of twist two, one can compute the second TMM in the TMD scheme, which agrees with the  $\MS$ scheme up to NLO. We emphasize that the expression in Eq.~(\ref{def:AS_X0}) holds for unpolarized, helicity, and transversity functions, with the relevant kernels and coefficient functions substituted accordingly.

\subsection{General scales}

A similar result (albeit with notably more intricate expressions) can be derived for TMDs assessed at the set of scales $\mu=\mu_{\text{OPE}}=\mu_{\text{TMD}}=\sqrt{\zeta}$. 

The second moment is determined by the transformation $\mathcal{G}$ defined in Eq.~(\ref{def:Gnplus1n}). For general scales, this integral cannot be reduced to the collinear matrix element $\mathbb{M}_{\mu\nu}$ in Eq.~(\ref{def:Mn-operator}). This is evident from the structure of the OPE, which takes the form
\begin{eqnarray}
\widetilde{F}(x,b;\mu_{\text{OPE}},\mu_{\text{TMD}},\zeta)&=&C(\mathbf{L}_{\mathcal{O}},\mathbf{L}_T,\pmb{{\ell}}_{T}) \otimes f
\\\nn &&
+
b^2 \(
\sum_k C^{(2)}(\mathbf{L}_{\mathcal{O}},\mathbf{L}_T,\pmb{{\ell}}_{T})\otimes f_k^{\text{tw4}}
-\mathcal{D}_2\pmb{{\ell}}_{T} 
C(\mathbf{L}_{\mathcal{O}},\mathbf{L}_T,\pmb{{\ell}}_{T}) \otimes f\)+\mathcal{O}(b^4),
\end{eqnarray}
where $f$ is the twist-two collinear PDF, $f_k^{\text{tw4}}$ represents a twist-four collinear PDF, the coefficient $C$ is the leading power coefficient function given in Eq.~(\ref{C-general}), the coefficient $C^{(2)}$ is the subleading power coefficient function, and $\mathcal{D}_2$ stands for the power correction to the Collins-Soper kernel. It is noteworthy that $\mathcal{D}_2$ is not necessarily suppressed by $\alpha_s$ (see, e.g., Ref.~\cite{Vladimirov:2020umg}). The matrix element $\mathbb{M}_{\mu\nu}$ is related to $f_k^{\text{tw4}}$. The transformation $\mathcal{M}_{\mu\nu}$, computed at general scales, contains an additional contribution $\sim \mathcal{D}_2$, and thus, cannot be unambiguously related to $\mathbb{M}_{\mu\nu}$.

However, for the  scales shown in Eq.~(\ref{mu=...=zeta}), the contribution $\sim \mathcal{D}_2$ vanishes, allowing for the determination of $\mathbb{M}_{\mu\nu}$ as outlined in \sec{2nd}. In this case, the expression for $\text{AS}[\mathcal{G}_{1,0}[F]]$ differs from Eq.~(\ref{def:AS_X0}). It  can be derived from Eq.~(\ref{C-general}), and it reads as
\begin{eqnarray}\nn
&&
\text{AS}[\mathcal{G}_{1,0}[F]](x,\mu,\mu,\mu^2)=
\Big\{\alpha_s\Big[P_1+\frac{\Gamma_0+\gamma_1}{2}\Big]
\\\nn &&\qquad
+\alpha_s^2\Big[
P_2+\(\overline{C}_1-P_1-\Gamma_0-\frac{\gamma_1}{2}\)\otimes\(P_1-\beta_0+\frac{\Gamma_0+\gamma_1}{2}\)
+\Gamma_0^2\frac{2\zeta_3-1}{4}+\frac{\Gamma_1+\gamma_2+2d^{(2,0)}}{2}
\Big]
\\\nn &&\qquad
+\alpha_s^3\Big[
P_3+\(\overline{C}_1-P_1-\Gamma_0-\frac{\gamma_1}{2}\)\otimes\(P_2-\beta_1+\frac{\Gamma_1+\gamma_2}{2}+d^{(2,0)}+\Gamma_0^2\frac{2\zeta_3-1}{4}\)
\\\nn &&\qquad \quad+
\(\overline{C}_2-P_2-\Gamma_1-\frac{\gamma_2}{2}-d^{(2,0)}-\Gamma_0\beta_0\frac{4\zeta_3}{3}+\Gamma_0^2\frac{4\zeta_3+3}{4}\)\otimes
\(P_1-2\beta_0+\frac{\Gamma_0+\gamma_1}{2}\)
\\\nn &&\qquad \quad -
\(\overline{C}_1-P_1+\Gamma_0\frac{4\zeta_3-3}{2}-\frac{\gamma_1}{2}\)
\otimes \(P_1-\beta_0+\Gamma_0+\frac{\gamma_1}{2}\)\otimes \(P_1-2\beta_0+\frac{\Gamma_0+\gamma_1}{2}\)
\\ \nn &&\qquad \quad+
\Gamma_0^3\frac{2-\zeta_3-3\zeta_5}{4}+\Gamma_0^2\beta_0\frac{4\zeta_3+1}{4}
-\Gamma_0\beta_0^2\frac{2\zeta_3}{3}
+\Gamma_0\Gamma_1\frac{2\zeta_3-1}{2}
\\ &&\qquad\qquad +2d^{(2,0)}\beta_0+\frac{\Gamma_2+\gamma_3}{2}+d^{(3,0)}\Big]+\mathcal{O}(\alpha_s^4)\Big\}\otimes f(x,\mu),
\label{eq:asymptotics}
\end{eqnarray}
where all terms with anomalous dimensions are accompanied by $\mathbf{1}$ that is omitted for simplicity.
Notice that several recursive combinations appear in \eqref{asymptotics}, this is a known feature of expansions at higher order of $\alpha_s$.

\subsection{Phenomenological examples}

\begin{figure}[t]
\centering
\includegraphics[width=1.\textwidth]{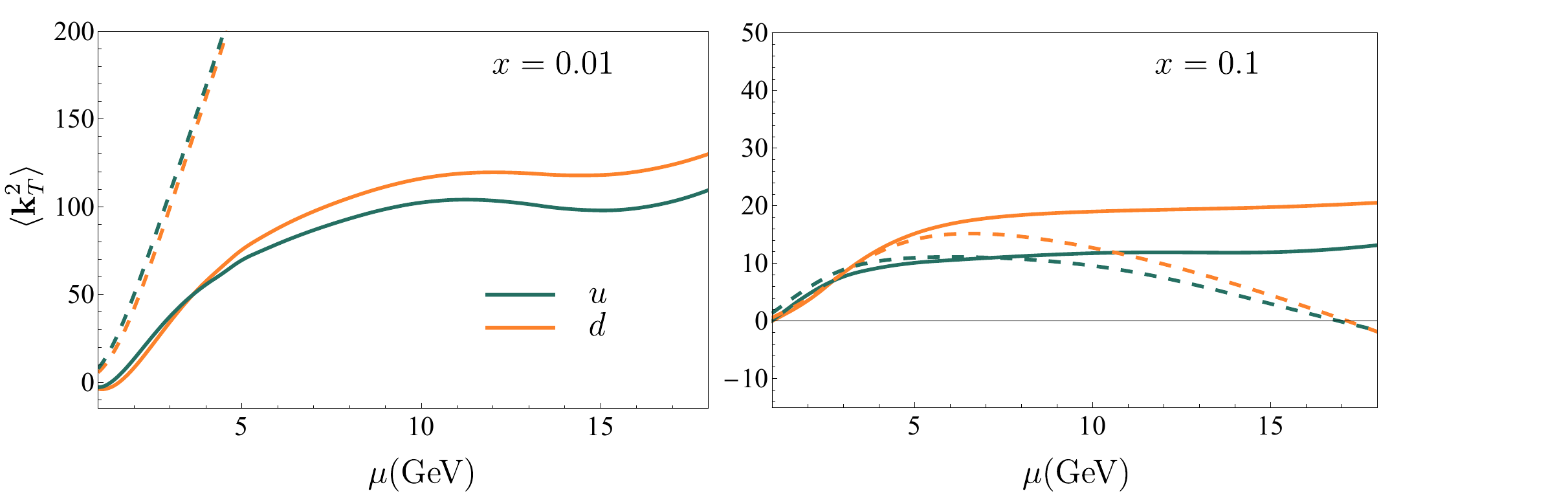}
\caption{Values of $\langle \vec k_T^2 \rangle$ computed for ART23 extraction of unpolarized TMD as a function of $\mu$. Solid and dashed orange lines are for $d$ quark, and solid and dashed blue lines are for $u$ quark. Dashed lines show the value of $\langle \vec k_T^2 \rangle$ without subtraction term.}
\label{fig:X0-vsMU}
\end{figure}

\begin{figure}[t]
\centering
\includegraphics[width=0.44\textwidth]{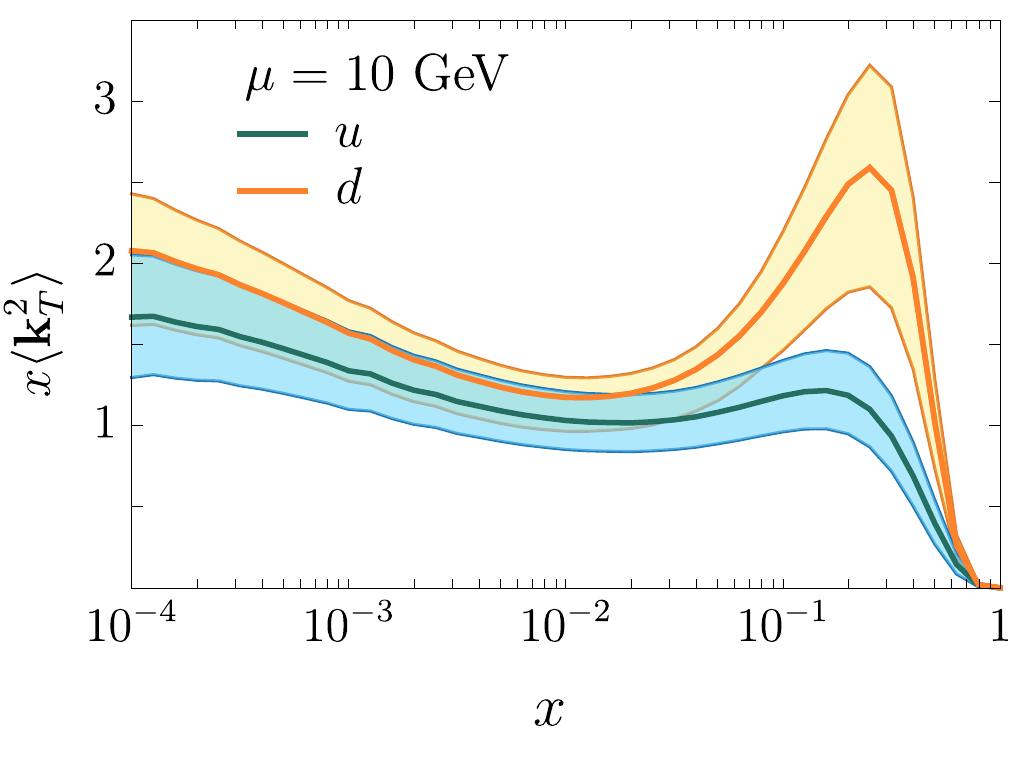}
~~
\includegraphics[width=0.44\textwidth]{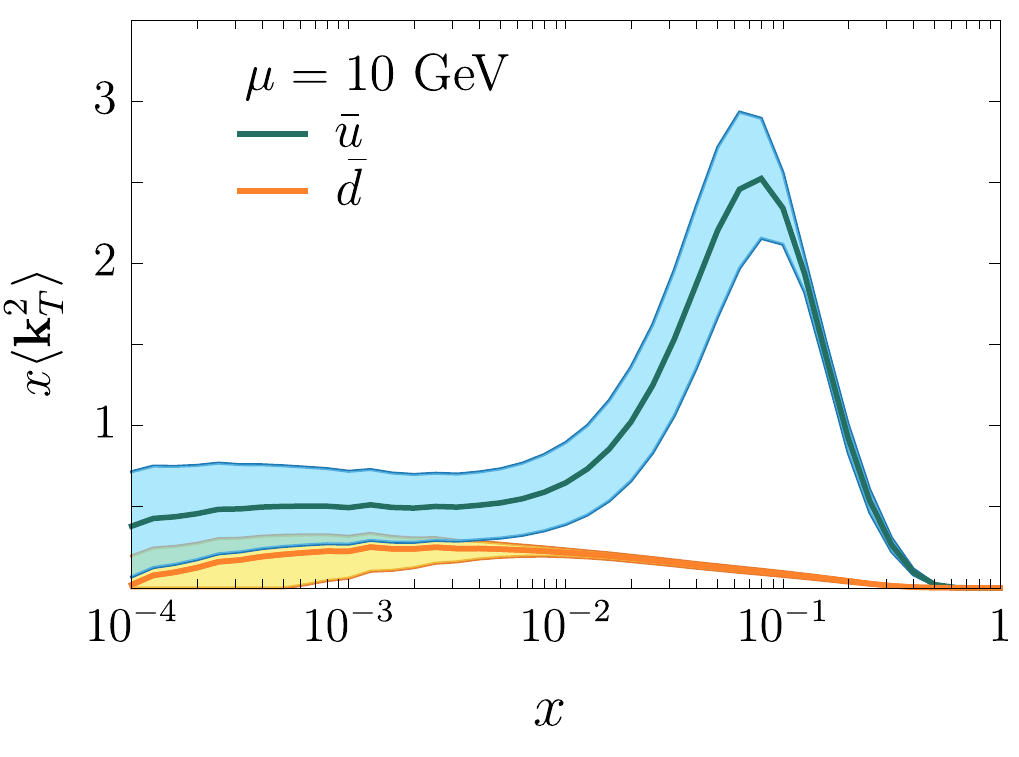}
\caption{Values of $x \langle \vec k_T^2 \rangle$ computed for ART23 extraction of unpolarized TMD as a function of $x$ at $\mu=10$ GeV. The two colors distinguish different flavors. The uncertainty band is evaluated from the uncertainty band of ART23 extraction.}
\label{fig:X0-vsX}
\end{figure}

The knowledge of the average width of TMDs can shed information on the dynamics of QCD. Reference \cite{Schweitzer:2012hh} using the ideas of the emergence of  $q\bar{q}$ condensate and model calculations predicted that the sea quark TMDs should be different and wider compared to the valence quark TMDs. An attempt to numerically study the widths was made in Ref.~\cite{Bacchetta:2022awv}, where the authors realized the difficulties of the naive application of integration of TMDs in $\vec k_T$ and used a ``renormalization" procedure consisting of calculating the result at a fixed $b$ instead of $b=0$. An alternative observable to the momentum space width, the width in $b$ space $\int d^2 {\bm b}\, b\, \tilde{f}_{1}(x, b;Q,Q^2)$  was introduced and studied in Ref.~\cite{Barry:2023qqh}. The widths of $k_T$ and $b$ space provide complementary information, but they cannot be unambiguously related to each other in a model-independent way.

A meaningful estimate of the width of distributions in the momentum space is via the second TMM, $\langle \vec k^2_T\rangle = -g_T^{\mu\nu}\mathcal{M}^{[\gamma^+]}_{\mu\nu} = 2 M^2 \overline{\mathcal{G}}_{1,0}[f_1]$.
An example of the evaluation of $\langle \vec k^2_T\rangle$ is presented in Fig.~\ref{fig:X0-vsMU}.  It confirms that the asymptotic term cancels the power growth of the $\mathcal{G}_{1,0}$ transformation.\footnote{
However, the cancellation is not very precise because we use $\text{AS}[\mathcal{G}_{1,0}]$ only at N$^3$LO. Therefore, at very large values of $\mu$ (we have used $\mu_{\text{max}}=100$ GeV in our studies), one observes a power growth in $\overline{\mathcal{G}}_{1,0}$. The terms that generate this growth are induced by higher perturbative order (N$^4$LO in the present example) logarithms in the evolution of $f$ that are not compensated by the coefficient function. Thus, the cancellation of the asymptotic term can be facilitated by a minor variation of the scale. We have found that for our computation, it is sufficient to change $\mu\to 1.014 \mu$ to eliminate the asymptotic term very precisely in a wide range of $\mu$.} The resulting line exhibits a general logarithmic behavior. In Fig. \ref{fig:X0-vsMU}, we do not present uncertainty bands because they are quite large, on the order of $25\%$-$35\%$ for $u$ and $d$ quarks (at $\mu=20$ GeV), and $50\%$-$60\%$ for $\bar u$ and $\bar d$ quarks.

We have also observed that at large $\mu$, the curve for $\text{AS}[\mathcal{G}_{1,0}]$ oscillates (the first sign of oscillation is seen in the left panel of Fig. \ref{fig:X0-vsMU}). This is a result of the number-flavor-variation scheme used in ART23. The change of $N_f$ in the coefficient function generates tiny discontinuities in $F(x,b)$ at certain values of $b$. In turn, they generate oscillations in the Hankel transform. The oscillations are small (see, e.g., Fig. \ref{fig:TMDtoPDF_fixedX}, where the same oscillations are present but barely visible). However, the subtraction of the asymptotic term amplifies the oscillations by a power factor and affects the entire procedure. This implies that the determination of second moments places more stringent requirements on the TMD model.

In Fig.~\ref{fig:X0-vsX}, we plot  $x \langle \vec k^2_T\rangle$ as a function of $x$ at $\mu=10$ GeV. Notice that one can see from Fig.~\ref{fig:X0-vsX} that the width resulting from ART23 extraction grows with the decrease of $x$. This growth is associated with the effects of the gluon shower that is characteristic of high energies, see for instance Ref.~\cite{Hautmann:2013tba} where the widening of the gluon transverse momentum dependent density was found. We also see that ART 23 has flavor dependence for widths, with antiquarks being narrower than quarks, except for $\bar u$ at $x\sim 0.1$, where it is similar to $u$ and $d$. 
If we consider the widths averaged with $x$,
\begin{eqnarray}\label{eq:xkt}
\langle x \vec k_T^2\rangle\ = 2 M^2 \int_0^1 dx \, x \,  \overline{\mathcal{G}}_{1,0}[f_1],
\end{eqnarray}
using ART23 extraction, we obtain
\begin{align} \label{eq:avkt2}
\nonumber
\langle x \vec k_T^2\rangle_{u}&=0.52 \pm 0.12 \; \text{GeV}^2,\qquad
\langle x \vec k_T^2\rangle_{d}=1.10 \pm 0.28 \; \text{GeV}^2,
\\ 
\langle x \vec k_T^2\rangle_{\bar u}&=0.42 \pm 0.06 \; \text{GeV}^2,\qquad
\langle x \vec k_T^2\rangle_{\bar d}=0.024 \pm 0.004 \; \text{GeV}^2.
\end{align}
 The inclusion of the $x$ weight in Eq.~(\ref{eq:xkt}) is required to facilitate the convergence of the integral at small $x$. While one might anticipate that the integrals $\langle \vec k_T^2\rangle$ for valence combinations would naturally converge, akin to the PDF case, this convergence should be explicitly incorporated into the fitting ansatz. In the case of ART23, the parameters governing the low $x$ behavior of quarks and antiquarks were treated independently, resulting in divergent integrals.  In the future, it will be interesting to consider parametrizations in which $\langle \vec k_T^2\rangle_{q}$ could be numerically estimated.

The numbers we obtain in Eqs.~(\ref{eq:avkt2}) are in a reasonable agreement with those found in the parton model extractions and flavor independent fits, for instance in Refs.~\cite{Anselmino:2005ea}, where it was found that $\langle \vec k_T^2\rangle \approx 0.25$ GeV$^2$. The parton model fit \cite{Anselmino:2013lza} of HERMES multiplicities in SIDIS resulted in $\langle \vec k_T^2\rangle =0.57\pm 0.08$ GeV$^2$. These numbers are similar to those found in the analyses with TMD evolution in Refs.~\cite{Bacchetta:2022awv,Bacchetta_2017}.

\section{Conclusions \label{sec:conclusions}}
 
In this paper, we explore the relationship between TMD distributions and collinear functions. We define TMMs, the weighted integrals of TMDs with a weight $\vec k_T^n$ and a momentum cutoff $|\vec k_T| < \mu$, such that the cutoff, $\mu$, becomes the scale at which the collinear functions are evaluated. We prove that the evolution of TMMs is the same DGLAP(-type) evolution equations as for the collinear quantities. The relation holds in the case of the $\zeta$ prescription or in the general case of all scales of TMDs being equal. We find it particularly convenient to utilize TMM with TMDs defined using the $\zeta$ prescription~\cite{Scimemi:2018xaf}, which allows one to use  scaleless nonperturbative TMD directly and simplifies analytic calculations. We prove that TMMs are equal to collinear distributions evaluated in some minimal subtraction schemes which we called a TMD scheme for TMDs in the $\zeta$ prescription and a TMD2 scheme for TMDs in a general TMD formalism. We demonstrate that the relation between TMD and a $\MS$ scheme, which is typical for collinear observables, is purely perturbative via a perturbatively calculable coefficient $Z$. Using this coefficient $Z$ the TMMs can be converted to the $\MS$ scheme.

In particular, we consider the first three TMMs that provide the most interesting phenomenological pieces of information.

The zeroth TMM is related  to collinear twist-2 PDFs (unpolarized, helicity, or tranversity, depending on the input TMD). We prove that this TMM obeys the same DGLAP evolution equation as collinear PDFs and we derive the coefficient $Z$ up to three loops. As a phenomenological example, we have performed an exhaustive  comparison of the zeroth TMM using unpolarized TMDs from a global QCD 
analysis of the data at N4LL (ART23) and compared the results to the corresponding collinear PDF MSHT20 obtained in the analysis of the data with NNLO precision.
We have shown that for $\mu\gtrsim 5$ GeV both the central values and the errors agree. 
The extraction of PDFs proceeds currently through the analysis of data from either inclusive processes or processes with high transverse momentum, for which a collinear description is appropriate. Nevertheless, the same PDF could be extracted also using data compatible with TMD factorization, through the procedure outlined in this work. This observation is important as it is a further step toward a simultaneous extraction of PDFs and TMDs.

The first TMMs involve TMDs like Sivers, Boer-Mulders functions, and worm-gear TMD functions. These TMMs provide information on the average transverse momentum shift of partons due to the correlation of the transverse motion and the spin and/or to the spin of the partons and that of the nucleon. The evolution of the first TMMs is the same in TMD  and $\overline{\rm MS}$ schemes up to ${\cal O}(\alpha_s^2)$. The first TMM in the case of the Sivers functions corresponds to a certain projection of the collinear twist-3 functions, the so-called Qiu-Sterman function. Using the current extraction of the Sivers functions we have provided an explicit calculation of the first TMMs. Moreover, utilizing the information on the quark Sivers functions we have estimated the average shift of gluons in a transversely polarized proton.

Finally, we have established that the second TMMs are related to the average widths of unpolarized, helicity, and transversity TMDs. The second TMMs have power divergence and need a subtraction to yield results that can be related to collinear quantities.  Therefore, a proper subtraction scheme is introduced, leading to subtracted averaged second TMMs that provide a quantitative estimate of collinear twist-4 operators.  We have provided a phenomenological example using ART23 extraction of TMDs.

We conclude by stating that establishing the relation between the 1D and 3D structure of the nucleon holds significant importance from theoretical, phenomenological, and experimental standpoints. This connection offers a unified approach to investigate the nucleon's structure, aiding in the development of comprehensive global QCD analyses, and could help to shape the experimental initiatives at current and prospective facilities, such as Jefferson Lab 12 and the future Electron-Ion Collider.

The results of this paper open new avenues for theoretical and phenomenological investigation of the three-dimensional and collinear hadron structures. We trust that usage of our results will facilitate phenomenological extractions of TMDs and collinear distributions and it will be useful in lattice QCD studies.

The supporting data for this paper, specifically the Z factors, are openly available from~\cite{github}.

\acknowledgments
We would like to thank C\'edric Lorc\'e for the correspondence about this paper. This work was partially supported by the National Science Foundation under Grants  No.~PHY-2012002, No.~PHY-2310031, and No.~PHY-2335114 (A.P.), and the U.S. Department of Energy Contract No.~DE-AC05-06OR23177, under which Jefferson Science Associates, LLC operates Jefferson Lab (A.P.).
A.V. is funded by the \textit{Atracci\'on de Talento Investigador} program of the Comunidad de Madrid (Spain) No. 2020-T1/TIC-20204 and \textit{Europa Excelencia} EUR2023-143460, MCIN/AEI/10.13039/501100011033/,  from Spanish Ministerio de Ciencias y Innovaci\'on. O. dR. is supported by the MIU (Ministerio de Universidades, Spain) fellowship FPU20/03110. 
This project is supported by the Spanish Ministerio de Ciencias y Innovaci\'on Grant No. PID2022-136510NB-C31 funded by MCIN/AEI/ 10.13039/501100011033. This project has received funding from the European Union Horizon 2020 research and innovation program under Grant Agreement No. 824093 (STRONG-2020).

\appendix
\setcounter{secnumdepth}{0}
\section{APPENDIX: PERURBATIVE EXPRESSIONS FOR $\zeta$ PRESCRIPTION}
\label{app:zeta-appendix}

In this Appendix, we collect the expressions for the computation of TMD in the  $\zeta$ prescription relevant to the present discussion. The  $\zeta$ prescription was developed in Refs. \cite{Vladimirov:2019bfa, Scimemi:2018xaf}, where detailed evaluations and additional expressions can be found.

\subsection{1. The optimal equipotential line in the perturbative regime}

The optimal equipotential line is defined by the differential equation Eq.~(\ref{zeta-line}) along with the specified boundary conditions Eq.~(\ref{def:saddle}). Note that this equation is generally nonperturbative due to the involvement of the Collins-Soper kernel $\mathcal{D}$, which is inherently nonperturbative. The solution, expressed in terms of a general $\mathcal{D}$, can be found in Ref.~\cite{Vladimirov:2019bfa}. 

The perturbative expression for the optimal equipotential line reads as \cite{Scimemi:2019cmh}
\begin{eqnarray}
\zeta_\mu^{\text{pert}}(b)=\frac{\mu}{b}2e^{-\gamma_E}e^{-v(b,\mu)},
\end{eqnarray}
where
\begin{eqnarray}
v(b,\mu)=\sum_{n=0}^\infty \alpha_s^n v_n(\mathbf{L}_\mu).
\end{eqnarray}
The coefficients $v_n$ up to NNLO are
\begin{eqnarray}
v_0(\mathbf{L}_\mu)&=&\frac{\gamma_1}{\Gamma_0},
\\
v_1(\mathbf{L}_\mu)&=&
\frac{\beta_0}{12}\mathbf{L}_\mu^2-\frac{\gamma_1\Gamma_1}{\Gamma_0^2}
+\frac{\gamma_2+d^{(2,0)}}{\Gamma_0},
\\
v_2(\mathbf{L}_\mu)&=&
\frac{\beta_0^2}{24}\mathbf{L}_\mu^3 +\(\frac{\beta_1}{12}+\frac{\beta_0\Gamma_1}{\Gamma_0}\)\mathbf{L}_\mu^2
+\(\frac{\beta_0\gamma_2}{2\Gamma_0}+\frac{4\beta_0d^{(2,0)}}{3\Gamma_0}-\frac{\beta_0\gamma_1\Gamma_1}{2\Gamma_0^2}\)\mathbf{L}_\mu
\\\nn &&+\frac{\gamma_1\Gamma_1^2}{\Gamma_0^3}-\frac{\gamma_1\Gamma_2+\gamma_2\Gamma_1+d^{(2,0)}\Gamma_1}{\Gamma_0^2}+\frac{\gamma_3+d^{(3,0)}}{\Gamma_0}.
\end{eqnarray}

\subsection{2. Small-$b$ coefficient function}
\label{app:zeta-coeff}

The coefficient functions for the small-$b$ OPE, as computed in Refs. \cite{Echevarria:2016scs, Luo:2019szz, Ebert:2020yqt, Luo:2020epw, Bacchetta:2013pqa, Gutierrez-Reyes:2017glx, Gutierrez-Reyes:2018iod, Gutierrez-Reyes:2019rug, Echevarria:2015usa, Luo:2019hmp, Luo:2019bmw, Ebert:2020qef, Scimemi:2019gge, Rein:2022odl}, are provided for the general TMD scale setting $(\mu,\zeta)$ at $\mu_{\text{TMD}}=\mu_{\text{OPE}}=\mu$. These coefficients exhibit a double-logarithmic form
\begin{eqnarray}\label{app:coeff}
\overline{C}(\mathbf{L}_{\mathcal{O}},\pmb{{\ell}}_T)=\sum_{n=0}^\infty \sum_{k=0}^{2n}\sum_{l=0}^n \alpha_s^n \mathbf{L}_{\mathcal{O}}^k \pmb{{\ell}}_T^l c^{(n,k,l)},
\end{eqnarray}
where $c^{(n,k,l)}$ are expressions involving collinear momentum fractions, and we denote the logarithms as
\begin{eqnarray}
\mathbf{L}_{\mathcal{O}}=\ln\(\frac{\mu_{\text{OPE}}^2 b^2}{4e^{-2\gamma_E}}\),
\qquad
\pmb{{\ell}}_T=\ln\(\frac{\mu_{\text{TMD}}^2}{\zeta}\).    
\end{eqnarray} 

The coefficients $c^{(n,k,l)}$ for $k\neq 0$ involve evolution kernels and anomalous dimensions, and, as previously done, we define $c^{(n',0,0)}=\overline{C}_{n'}$ with $n'<n$ [see Appendix D in Ref.~\cite{Echevarria:2016scs} or expression Eq.~(\ref{C-general})].  In the  $\zeta$ prescription, the double-logarithm terms vanish, and Eq.~(\ref{app:coeff}) turns to

\begin{eqnarray}\label{app:zetacoeff}
\widetilde{C}(\mathbf{L}_{\mathcal{O}})=\sum_{n=0}^\infty \sum_{k}^{n} \alpha_s^n \mathbf{L}_{\mathcal{O}}^k  C^{(n,k)}.
\end{eqnarray}

Similar to the general case, the coefficients $C^{(n,0)}=C_n$ are the unique part of this expression, as the rest of the terms are expressed via evolution kernels [see an example in Eq.~(\ref{coef-zeta})].

The relation between $C_n$ and $\overline{C}_n$ is as follows
\begin{eqnarray}
C_{1}&=&\overline{C}_{1},
\nn \\
C_{2}&=&\overline{C}_2+\frac{d^{(2,0)} \gamma_1}{\Gamma_0},
\label{eq:ccbar}\\
C_{3}&=&\overline{C}_3+\frac{d^{(2,0)} \gamma_1}{\Gamma_0}\overline{C}_1 
+
\(\frac{(d^{(2,0)})^2+d^{(3,0)}\gamma_1+d^{(2,0)}\gamma_2}{\Gamma_0}-\frac{d^{(2,0)}\gamma_1\Gamma_1}{\Gamma_0^2}\), \nn
\end{eqnarray}
where coefficients $\Gamma$, $\gamma$, and $d$ are defined in Eqs.~(\ref{eq:Gamma}) and (\ref{eq:gammaV}) and Eq.~(\ref{eq:cs-expansion}). The coefficient functions are matrices in flavor space, and the flavor indices in the expressions above are contracted in the natural way. The two-loop $\bar C_2$ matrix is reported in Eqs.~(7.3)-(7.8) of Ref.~\cite{Echevarria:2016scs} and in the Mathematica auxiliary files of Ref.~\cite{Echevarria:2016scs} and the three-loop $\bar C_3$ can be found in Refs.~\cite{Luo:2019szz,Ebert:2020yqt}.

Expressions for coefficients $ C_1$  read as, see Ref.~\cite{Echevarria:2016scs} and references therein:
\begin{eqnarray}\label{eq:matrix coef zeta}
    C^{qq}_1(x)&=&C_F\(2(1-x)-\frac{\pi^2}{6}\delta(1-x)\),\nn \\
    C^{qg}_1(x)&=&4 T_R(1-x)x,\nn  \\
    C^{gq}_1(x)&=&2C_Fx,\\
    C^{gg}_1(x)&=&-C_A\frac{\pi^2}{6}\delta(1-x),\nn\\
    C^{\bar{q}q}_1(x)&=&C^{q'q}_1(x)=0,\nn
\end{eqnarray}
where $\bar{q}$ denotes the antiquark of the quark $q$, and $q'$ indicates the other possible flavors of quarks different form $q$ and $\bar{q}$. 
%might be too lengthy already to write here for example:
%\begin{eqnarray}
%    \nn C^{qq}_2(x)&=&\frac{C_F}{648 (1-x)}\Bigg(9 C_F \Big(24 (12 (x^2+1) \text{Li}_3(1-x)-60 (x^2+1) \text{Li}_3(x)+36 (x^2+1) \text{Li}_2(x) \ln (x)\\
%   \nn &&+12 (x^2+1) \text{Li}_2(x) \ln (1-x)-24 (x-1)^2 \text{Li}_2(x)+60 (x^2+1) \zeta (3)-x^2 \ln ^3(x)+6 (x^2+1) \ln (1-x) \ln ^2(x)\\
%    \nn&&+18 (x^2+1) \ln ^2(1-x) \ln (x)-2 ((\pi ^2-3) x^2+3 x+\pi ^2) \ln (1-x)+3 (\pi ^2-22) (x-1)^2+\ln ^3(x)\\
%    \nn&&-(6 (x-1) x-9) \ln ^2(x)+6 ((16 x-13) x+5) \ln (x)-36 (x-1)^2 \ln (1-x) \ln (x))-\pi ^4 (x-1) \delta(1-x)\Big)\\
%    \nn&&-2 C_A \Big(-48 (202-189 \zeta (3)) ((x-1) \left(\frac{1}{1-x}\right)_++1)+(x-1) (4032 \zeta (3)+18 \pi ^4-603 \pi ^2-2416) \delta(1-x)\\
%    \nn&&+6 (216 (x^2+1) \text{Li}_3(1-x)-432 (x^2+1) \text{Li}_3(x)+216 (x^2+1) \text{Li}_2(x) \ln (x)+216 (x^2+1) \text{Li}_2(x) \ln (1-x)\\
%    \nn&&-216 (x-1)^2 \text{Li}_2(x)-324 (x^2+1) \zeta (3)+18 x^2 \ln ^3(x)+216 (x^2+1) \ln ^2(1-x) \ln (x)\\
%    \nn&&+12 (83 x^2-36 x+29) \ln (x)-36 ((\pi ^2-3) x^2+3 x+\pi ^2) \ln (1-x)+54 \pi ^2 (x-1)^2+16 ((x+99) x+1)\\
%    \nn&&+18 \ln ^3(x)+9 (11-(x-12) x) \ln ^2(x)-216 (x-1)^2 \ln (1-x) \ln (x))\Big)\\
%    \nn&&+\frac{1}{81} T_R n_f\Big((252 \zeta (3)+45 \pi ^2+176) \delta(1-x)-\frac{6}{x-1} (-112 (x-1) \left(\frac{1}{1-x}\right)_++38 x^2\\
%    \nn&&+9 (x^2+1) \ln ^2(x)+30 (x^2+1) \ln (x)+36 x-74)\Big)\Bigg)
%\end{eqnarray}}

%\oscar{In Ref.~\cite{Ebert:2020yqt} they even give mathematica files for these coefficients but we can do one ourselves for  $\zeta$ prescription, what do you think? }
We report here the known $d^{(i,0)}$  (see~\cite{Becher:2010tm,Echevarria:2015byo,Li:2016ctv,Vladimirov:2016dll,Vladimirov:2017ksc}):
\begin{align}
d^{(1,0)}&=0,\qquad d^{(2,0)}=C_A\(\frac{404}{27}-14 \zeta_3\)-\frac{56}{27}N_f,
\\\nn
d^{(3,0)}&=C_A^2\(\frac{297029}{1458}-\frac{3196}{81}\zeta_2-\frac{6164}{27}\zeta_3-\frac{77}{3}\zeta_4+\frac{88}{3}\zeta_2\zeta_3+96\zeta_5\)
\\\nn&+C_AN_f\(-\frac{31313}{729}+\frac{412}{81}\zeta_2+\frac{452}{27}\zeta_3-\frac{10}{3}\zeta_4\)
\\\nn&+C_FN_f\(-\frac{1711}{54}+\frac{152}{9}\zeta_3+8\zeta_4\)+N_f^2\(\frac{928}{729}+\frac{16}{9}\zeta_3\).
\end{align}

To compute the $Z$ factor between the TMD scheme and the  $\MS$ scheme, Eq.~(\ref{Z:TMD->MS}), we need to compute various convolutions, such as
\begin{equation}\label{eq:ConvolutionC1C1}
    (C_1\otimes C_1)^{ff'}(x)=\sum_{i}\int_x^1\frac{dy}{y}C_1^{fi}(y)C_1^{ if'}\left(\frac{x}{y}\right)
\end{equation}
The explicit results are
\begin{align}
\left(C_1\otimes C_1\right)^{qq} &=  C_F^2 \left(\frac{\pi ^4}{36} \delta(1-x)+\left(8+\frac{2\pi ^2}{3}\right) (x-1)-4 (x+1) \log (x)\right)-8 C_F T_R \Big(x (1-x)+x\log (x)\Big)\; ,\\
\left(C_1\otimes C_1\right)^{qg} &=C_F T_R \Big(4 (1-x^2)+8 x \log (x)-\frac{2\pi^2}{3}(1-x) x \Big) -\frac{2\pi^2}{3}C_A T_R(1-x) x \; , \\
\left(C_1\otimes C_1\right)^{gq} &= C_F^2\left(2 (1-x)^2-\frac{\pi ^2}{3}x\right) -\frac{\pi ^2}{3}C_F C_A x \; , \\
\left(C_1\otimes C_1\right)^{gg} &=\frac{\pi ^4}{36}  C_A^2 \delta(1-x)-16 C_F N_f T_R \Big(x (1-x)+x\log (x)\Big)\; , \\
\left(C_1\otimes C_1\right)^{\bar q q} &=\left(C_1\otimes C_1\right)^{ q'q}= 0 \; .
\end{align}

The same expressions are to be used for the TMD2 scheme Z factor in Eq.~(\ref{Z:TMD2->}) where a $\bar C_1\otimes \bar C_1$ convolution appears, as at NLO they are the same, $\bar C_1=  C_1$.

%\bibliography{bibFILE}

\end{document}